# Energy Circuit-based Integrated Energy Management System: Theory, Implementation, and Application

Binbin Chen, *Student Member, IEEE*, Qinglai Guo, *Senior Member, IEEE*, Guanxiong Yin, *Student Member, IEEE*, Bin Wang, Zhaoguang Pan, *Member, IEEE*, Yuwei Chen, Wenchuan Wu, *Fellow, IEEE*, Hongbin Sun, *Fellow, IEEE*

*Abstract*—Integrated energy systems (IESs), in which various energy flows are interconnected and coordinated to release potential flexibility for more efficient and secure operation, have drawn increasing attention in recent years. In this article, an integrated energy management system (IEMS) that performs online analysis and optimization on coupling energy flows in an IES is comprehensively introduced. From the theory perspective, an energy circuit method (ECM) that models natural gas networks and heating networks in the frequency domain is discussed. This method extends the electric circuit modeling of power systems to IESs and enables the IEMS to manage large-scale IESs. From the implementation perspective, the architecture design and function development of the IEMS are presented. Tutorial examples with illustrative case studies are provided to demonstrate its functions of dynamic state estimation, energy flow analysis, security assessment and control, and optimal energy flow. From the application perspective, real-world engineering demonstrations that apply IEMSs in managing building-scale, park-scale, and city-scale IESs are reported. The economic and environmental benefits obtained in these demonstration projects indicate that the IEMS has broad application prospects for a low/zero-carbon future energy system.

## I. Introduction

### A. IES: the Physical Kernel of Energy Internet

The "Energy Internet" is a new ecosystem that integrates energy systems (the physical layer) and the Internet (the cyber layer) to realize better openness and interconnection in energy production, transmission, and consumption [1]. Different from traditional energy systems, in which various energy sectors are managed separately [2], a key feature of the Energy Internet is that electricity networks, heating networks, natural gas networks, and other energy networks are operated and managed coordinately to form so-called integrated energy systems (IESs) [3] or multienergy systems (MESs) [4]. This feature is illustrated in Fig. 1.

It is well recognized that IESs can release potential flexibility through shifting across different energy sectors [5], which is reflected in (1) better energy supply reliability through multienergy complementation and switching [6]-[7], (2) higher operational efficiency through the optimization of the multienergy infrastructure configuration [8] and energy cascade utilization [9], and (3) more renewable energy accommodation through the conversion of superfluous power into other energy sectors such as natural gas [10] and heat [11], which are convenient and inexpensive in terms of storage. O'Malley *et al.* evaluated the IES as a potential pathway to a low/zero-carbon future energy system [12].

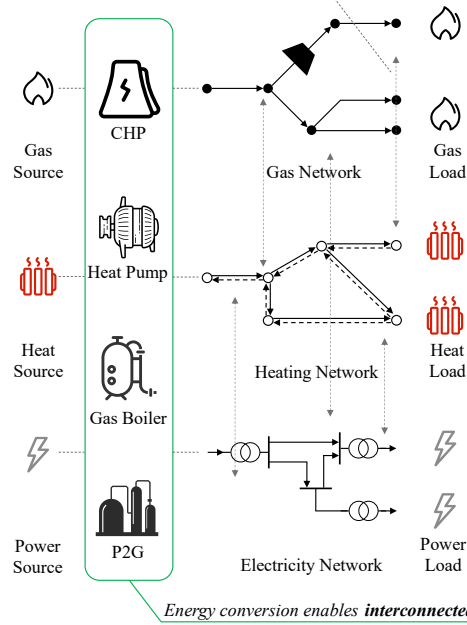

Fig. 1. Integrated energy systems realize interconnected operation and coordinated management of various energy sectors.

However, multienergy interconnection also has drawbacks in that the deep coupling between energy networks complicates IESs. Consequently, the control and management of IESs becomes more difficult, and risks of cascading accidents across energy sectors arise. Some blackouts in electricity networks due

Binbin Chen, Qinglai Guo, Guanxiong Yin, Bin Wang, Zhaoguang Pan, Yuwei Chen, and Wenchuan Wu are with the State Key Laboratory of Power Systems, Dept. Electrical Engineering, Tsinghua University, Beijing 100084, China (email: cbb18@mails.tsinghua.edu.cn; guoqinglai@tsinghua.edu.cn; yin_guanxiong@qq.com; wb1984@tsinghua.edu.cn; panzg09@163.com; 18811362415@163.com; wuwench@tsinghua.edu.cn).

Hongbin Sun is with the State Key Laboratory of Power Systems, Dept. Electrical Engineering, Tsinghua University, Beijing 100084, China, and also with the Taiyuan University of Technology, Taiyuan 030024, China (email: shb@tsinghua.edu.cn).

This work was jointly supported by the National Key Research and Development Program (Grant. 2020YFE0200400) and the National Natural Science Foundation of China (Grant. U2066206). *(Corresponding author: Hongbin Sun.)*

to failures in natural gas networks that occurred in Texas [13], California [14], and Taiwan [15] have raised concerns.

*B. IEMS: the Central Nervous System of IESs*

To ensure the secure and economic operation of IESs, an integrated energy management system (IEMS) that plays a central role in managing multiple energy flows in an IES has been developed.

The precursor of the IEMS can be traced back to the energy management system (EMS), which is regarded as the central nervous system for electricity network operation [16]-[18]. Although the management objects have been expanded from power flow to more energy flows, such as heat and natural gas, IEMSs provide essential functions similar to those of EMSs. Supervisory control and data acquisition (SCADA), which manages real-time asynchronized measurement data from different energy networks, is employed to monitor IESs [19]. These measurement data are further processed by state estimation to perceive IES states that are more accurate and consistent across different energy sectors [20]. Incorporating the estimated data with the forecasted data from load prediction that considers the correlation of multienergy loads [21], energy flow analysis and security assessment are performed to detect potential security risks, especially the risks of cascading failures in the context of IESs [22]-[23]. Moreover, optimal energy flow that exploits the advantage of multienergy complementation is calculated to coordinate controllable resources in IESs for the purpose of more efficient operation [24].

Although an IEMS has functions similar to those of an EMS, its development and implementation is not a simple replication of the latter due to the large differences in the physical properties of the energy flows managed by the two systems. The transient processes in electricity networks usually have a time scale of milliseconds, which means that the transient processes can be omitted in the normal operation of electricity networks. Thus, the EMS deals with a steady-state model of electricity networks which is described by algebraic equations. However, the transient processes in other energy networks of IESs, such as natural gas networks and heating networks, can last minutes to hours [25]-[27] so that they cannot be omitted in the normal operation of IESs; otherwise, it raises security and efficiency issues, which are clarified in [28] and [29], respectively. As a consequence, a dynamic IES model, which is described by both partial differential equations (PDEs) and algebraic equations, is addressed by the IEMS.

From the EMS to the IEMS, different mathematical models not only bring about differences in solving techniques but also lead to the change in the analysis manner. Considering that the current state is uniquely determined by the current boundary condition in a steady-state model, one section[1] is sufficient to describe the response of an electricity network with a certain excitation. Thus, EMSs are mainly designed in a section-based manner. In contrast, the current state is jointly determined by the historical state and the current boundary condition in a dynamic model so that one period[2] with enough length is needed to completely reflect the response of an IES with a certain excitation. This decides IEMSs should be designed in a period-based manner. For instance, the function of energy flow analysis gives the system states in the next several hours rather than the current state only, and the function of security assessment gives not only whether the system is secure but also when the system is insecure.

*C. Works and Contributions*

Currently, the mainstream solving techniques supporting IEMS functions rely on the finite difference method (FDM) to approximate derivatives with differences, which thereby converts intractable PDEs into algebraic equations. However, this method encounters the following issues in practice: (1) The discretization in both spatial and temporal dimensions introduces numerous mesh points [30], which leads to heavy computational burden and limits applicable IES scales. (2) Inappropriate spatial and temporal step lengths may cause divergence or oscillation of computation, which degrades the reliability of IEMSs. (3) The comprehensive initial conditions, such as the initial states inside pipelines, required by the FDM are difficult to obtain due to insufficient measurements in natural gas networks and heating networks. In addition, extant studies mostly focus on a certain function of IEMSs, such as broadly studied energy flow analysis and optimal energy flow, but few are carried out from a system perspective or with engineering practice involved.

To close these gaps, related works from the perspectives of theory, implementation, and application are introduced in this article, which correspond to the following threefold contributions:

1) Inspired by the electric circuit modeling of electricity networks, an energy circuit method (ECM) that models natural gas networks and heating networks in the frequency domain is introduced. Compared with extant time-domain modeling methods, the ECM produces fewer variables and equations/constraints, which yields higher computational efficiency. Moreover, this method is free of step lengths and does not require initial states inside pipelines.
2) Extensible software architecture and function architecture are designed for IEMSs. Then, based on the ECM, advanced applications of IEMSs, including dynamic state estimation, energy flow analysis, security assessment and control, and optimal energy flow, are developed in a period-based manner. Tutorial examples with illustrative case studies are provided to demonstrate the functions of IEMSs.
3) In recent years, our implemented IEMSs have been successively deployed in several industrial parks and cities for engineering demonstrations, which has brought about significant economic and environmental advantages. The project background, on-site deployment, and overall benefits of engineering demonstrations on IESs with different scales are reported and summarized for the promotion of IEMSs.

---

[1] A "section" refers to the state variables of the whole system at a certain moment. It can be interpreted as a snapshot of the system.

[2] A "period" refers to the state variables of the whole system over a period of time. It can be interpreted as multiple "sections" with enough small intervals.

The remainder of this article is organized as follows. Section II introduces the theory of energy circuit modeling. In Section III, the implementation of IEMSs in architecture design and function development is presented. In Section IV, engineering demonstrations of IEMSs in China are reported. Finally, Section V concludes this article and presents an outlook on future IEMSs.

## II. THEORY: ENERGY CIRCUIT MODELING

Regarding the organization of this section, the related works on IES modeling are first reviewed in subsection A. Then, analogous to the electric circuit modeling for transmission lines, the energy circuit modeling for natural gas pipelines and heating pipelines is briefly introduced in subsection B. Finally, by combining branch equations and topology equations, the energy circuit models of natural gas networks and heating networks are derived in subsection C.

### A. Related Works

Modeling is a perpetual topic in many studies since it constitutes the basis of analysis and optimization, so as it is in the research of IESs.

The pioneering work of modeling an IES began in energy hub (EH) research [31], which is an end-to-end model using a coupling matrix to describe the multienergy conversion and delivery from supplies to consumption. This model was first proposed by Andersson *et al.* in 2007 [32]. Next, techniques for coupling-matrix computation in EH models were proposed in [33] and [34], making EH models practical. Because an EH model does not model transmission lines/pipelines and the corresponding dynamic processes, the security concerns and flexibility resources inside multienergy networks are ignored, which makes the model more suitable for small-scale IESs.

Since 2016, IES models involving network characteristics have proliferated, such as those in [35]-[37]. These models contain simultaneous equations for electricity networks, natural gas networks, heating networks, and coupling devices in a steady state, which successfully consider security concerns and incorporate tractable calculations. However, the flexibility from the large inertia of natural gas networks and heating networks, as modeled and assessed in [38] and [39], is still omitted by the steady-state models. Replacing the steady-state equations of natural gas networks and heating networks with the dynamic PDEs after FDM discretization [40]-[41] can solve this issue theoretically but not practically for large-scale IESs due to the massive computational burden. The above modeling techniques are compared and summarized in [42].

Another dynamic modeling approach of IESs is developed from an electric circuit-analog perspective. This is not a totally novel idea: Robertson and Gross designed an electric circuit for the dynamic simulation of transient heat flow in pipelines as early as 1958 [43]. In recent years, this idea has been reapplied in the digital simulation of IESs. In [44], a circuit-analog model for steady-state natural gas networks is proposed for the simultaneous analysis of electricity networks and natural gas networks in one environment. Similar works for steady-state heating networks are implemented in [45] and [46]. Considering dynamic characteristics, time-domain circuit models with distributed parameters are derived for natural gas networks and heating networks in [47] and [48], respectively, which enable the reuse of electromagnetic transient programs (EMTPs) for IES simulation. However, these two works, as well as another similar work in [49], still rely on temporal and spatial discretization for model solving. In contrast, the original PDEs describing dynamic processes in IESs are converted into ordinary differential equations (ODEs) by circuit analogy in [50] and algebraic equations further by the Laplace transform in [51] and [52]. Similar circuit models that use the Laplace transform for algebraization are adopted in [53] and [54] to formulate optimal energy flow problems of IESs. As reported in these works, such circuit models in the complex frequency domain contain fewer variables due to avoiding finite difference discretization and thereby significantly relieve computational burden. However, these models involve intractable symbolic calculation about the Laplacian "*s*", which requires reduced-order approximation at the cost of accuracy, as concluded in [55].

The proposed energy circuit model is a new circuit-analog model for IESs that fully draws on the deduction of electric circuit modeling from the time domain to the frequency domain and from distributed parameters to lump parameters. This model accurately reflects the dynamic processes in IESs and is tractable enough for calculation.

TABLE I
CIRCUIT ANALOGY: FROM ELECTRIC CIRCUIT TO FLUID CIRCUIT AND THERMAL CIRCUIT

| Energy Circuits | Electric Circuit | Fluid Circuit | Thermal Circuit |
|---|---|---|---|
| Physical Object | transmission lines in electricity networks | natural gas pipelines in natural gas networks | heating pipelines in heating networks |
| Potential $\mathcal{P}$ | voltage $U$ | pressure $p$ | temperature $T$ |
| Flow $\mathcal{F}$ | current $I$ | gas flow $m$ | heat flow $h$ |
| Resistance *resists flow* | $R_e: U = R_e I$ | $R_f: p = R_f m$ | $R_t: T = R_t h$ |
| Conductance *resists potential* | $G_e: I = G_e U$ | $G_f: m = G_f p$ | $G_t: h = G_t T$ |
| Inductance *resists flow change* | $L_e: U = L_e \mathrm{d}I/\mathrm{d}t$ | $L_f: p = L_f \mathrm{d}m/\mathrm{d}t$ | $L_t: T = L_t \mathrm{d}h/\mathrm{d}t$ |
| Capacitance *resists potential change* | $C_e: I = C_e \mathrm{d}U/\mathrm{d}t$ | $C_f: m = C_f \mathrm{d}p/\mathrm{d}t$ | $C_t: h = C_t \mathrm{d}T/\mathrm{d}t$ |

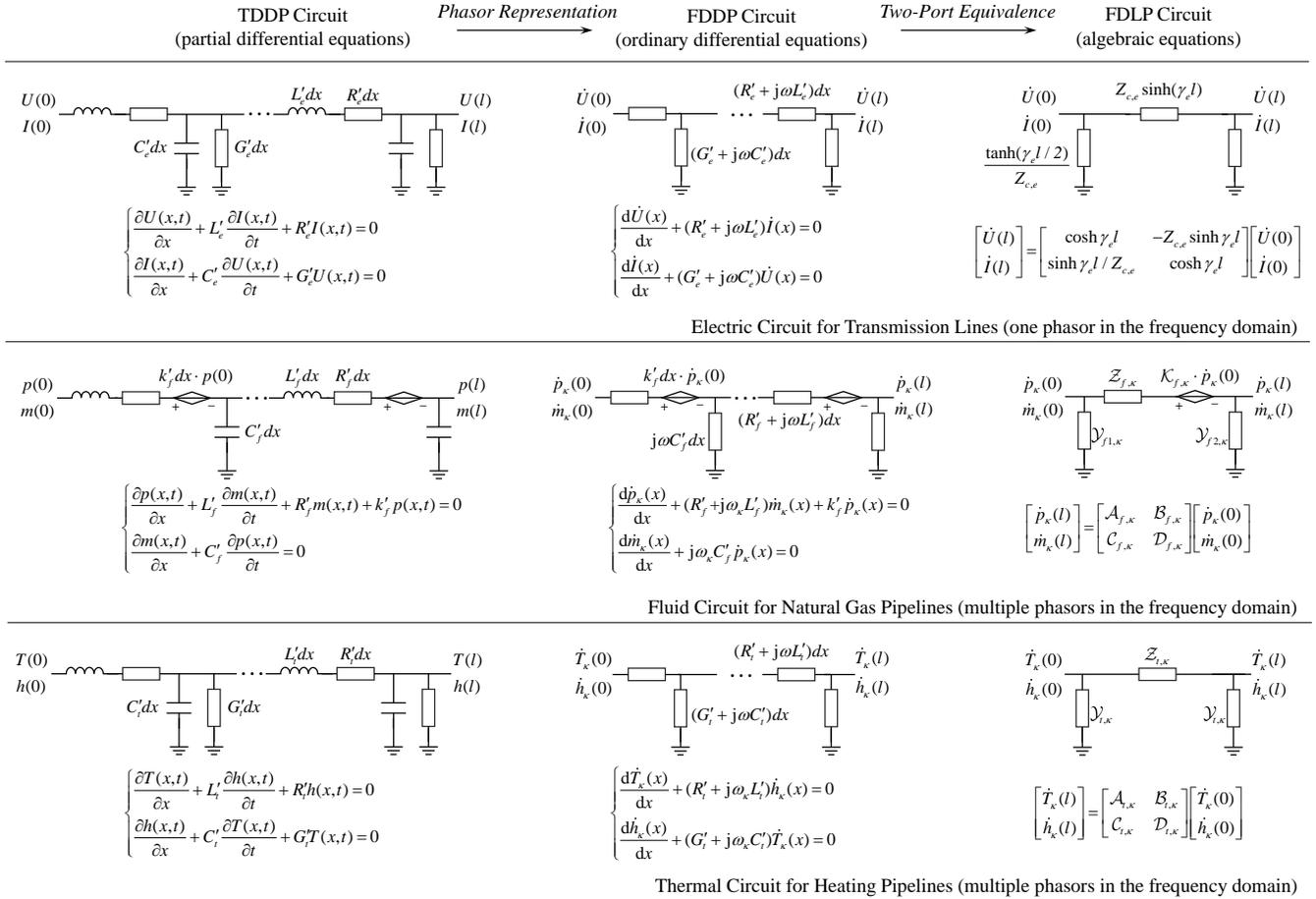

Fig. 2. Energy circuit modeling: from time domain to frequency domain and from distributed parameters to lump parameters on physics, and from partial differential equations to algebraic equations on mathematics.

### B. Energy Circuit Modeling for Pipelines

The voltage and current along AC transmission lines satisfy the telegrapher's equations [56], which are PDEs corresponding to a time-domain-distributed-parameter (TDDP) electric circuit model. Using the phasor method that converts a sine wave in the time domain into a complex vector to eliminate the temporal derivative [57], these PDEs are simplified to ODEs, which correspond to a frequency-domain-distributed-parameter (FDDP) electric circuit model. Further solving these ODEs yields an algebraic relationship between the voltage and current at both ends, which corresponds to a frequency-domain-lump-parameter (FDLP) electric circuit model, i.e., the famous pi-equivalent electric circuit [58]. Such a two-step transformation has brought significant computational convenience for power system analysis.

Through circuit analogy in TABLE I, the PDEs that describe the relationships of pressure and gas flow in natural gas pipelines, as in (1), and that describe the relationships of temperature and heat flow in heating pipelines, as in (2), can be reorganized into the form of the telegrapher's equations, which thereby enables the two-step transformation in electric circuit modeling for transmission lines to derive computation-efficient energy circuit models for natural gas pipelines and heating pipelines, as illustrated in Fig. 2.

$$\begin{cases} A\dfrac{\partial p}{\partial t} + c^2\dfrac{\partial m}{\partial x} = 0 \\ \dfrac{1}{A}\dfrac{\partial m}{\partial t} + \dfrac{\partial p}{\partial x} + \dfrac{\lambda \rho v^2}{2D} + \rho g \sin\alpha = 0 \\ m = \rho v A \\ p = c^2 \rho \end{cases} \quad (1)$$

where constants $A$, $D$, $\lambda$, $\alpha$, $g$, and $c$ are the cross-sectional area, inner diameter, friction factor, dip angle, gravitational acceleration, and sonic speed, respectively; variables $p$, $m$, $v$, and $\rho$ are the pressure, mass flow rate, velocity, and density, respectively.

$$\begin{cases} c_p \rho A \dfrac{\partial T}{\partial t} + c_p m \dfrac{\partial T}{\partial x} + \mu T = 0 \\ h = c_p m T \end{cases} \quad (2)$$

where constants $c_p$ and $\mu$ are the specific heat capacity and heat dissipation coefficient, respectively, and variables $T$ and $h$ are the relative temperature and heat flow, respectively. In this article, only heating networks operated with the quality regulation mode are considered, in which water flow $m$ of pipelines is a known constant optimized in advance for typical operating conditions and outlet temperature of heating sources is adjusted to follow heat load changes [59]. This regulation mode is broadly adopted for heating networks in Nordic countries, Russia, and north China [60].

More details about the circuit analogy in TABLE I and the energy circuit modeling in Fig. 2 are clarified as follows.

### Step 1) Circuit Analogy

As summarized in TABLE I, the state variables of different energy flows are categorized into "potential" $\mathcal{P}$ and "flow" $\mathcal{F}$: the potential variables include voltage $U$ in electricity networks, pressure $p$ in natural gas networks, and temperature $T$ in heating networks; the flow variables include current $I$ in electricity networks, gas flow $m$ in natural gas networks, and heat flow $h$ in heating networks. Based on this categorization, resistance $R$ that resists flow, conductance $G$ that resists potential, inductance $L$ that resists flow changes, and capacitance $C$ that resists potential changes are generalized from the electric circuit (with subscript $c$) to the fluid circuit[3] (with subscript $f$) and thermal circuit (with subscript $t$). By extracting these circuit components, the PDEs of natural gas pipelines and heating pipelines in (1) and (2) are reorganized:

$$\begin{cases} \dfrac{\partial p(x,t)}{\partial x} + L'_f \dfrac{\partial m(x,t)}{\partial t} + R'_f m(x,t) + k'_f p(x,t) = 0 \\ \dfrac{\partial m(x,t)}{\partial x} + C'_f \dfrac{\partial p(x,t)}{\partial t} = 0 \end{cases} \quad (3)$$

$$\begin{cases} \dfrac{\partial T(x,t)}{\partial x} + L'_t \dfrac{\partial h(x,t)}{\partial t} + R'_t h(x,t) = 0 \\ \dfrac{\partial h(x,t)}{\partial x} + C'_t \dfrac{\partial T(x,t)}{\partial t} + G'_t T(x,t) = 0 \end{cases} \quad (4)$$

These two PDEs give the TDDP circuits of natural gas pipelines and heating pipelines, as shown in the first column of Fig. 2. The expressions of the involved circuit component parameters, including $R'_f$, $L'_f$, $C'_f$, $k'_f$, $R'_t$, $G'_t$, $L'_t$, and $C'_t$, are provided in Appendix A.

### Step 2) Fourier Transform and Phasor Representation

To enable the phasor method, the Fourier transform is employed to decompose nonsinusoidal excitations in natural gas networks and heating networks into several sinusoidal components with different frequencies. The desired response is exactly the time-domain superposition of responses for each frequency component, as visualized in Fig. 3.

For $\kappa$-th frequency component whose angular frequency is $\omega_\kappa$, the TDDP equations in (3) and (4) are rewritten using the phasor representation:

---

[3] Although the fluid circuit in this article is derived based on the natural gas network equations, it is applicable to other fluid networks such as water

$$\begin{cases} \dfrac{d\dot{p}_\kappa(x)}{dx} + (R'_f + j\omega_\kappa L'_f)\dot{m}_\kappa(x) + k'_f \dot{p}_\kappa(x) = 0 \\ \dfrac{d\dot{m}_\kappa(x)}{dx} + j\omega_\kappa C'_f \dot{p}_\kappa(x) = 0 \end{cases} \quad (5)$$

$$\begin{cases} \dfrac{d\dot{T}_\kappa(x)}{dx} + (R'_t + j\omega_\kappa L'_t)\dot{h}_\kappa(x) = 0 \\ \dfrac{d\dot{h}_\kappa(x)}{dx} + (G'_t + j\omega_\kappa C'_t)\dot{T}_\kappa(x) = 0 \end{cases} \quad (6)$$

These two ODEs give the FDDP circuits of natural gas pipelines and heating pipelines, as shown in the second column of Fig. 2. Different from the electric circuit, there are a series of FDDP circuits with different frequencies corresponding to the TDDP circuit for the fluid circuit and thermal circuit.

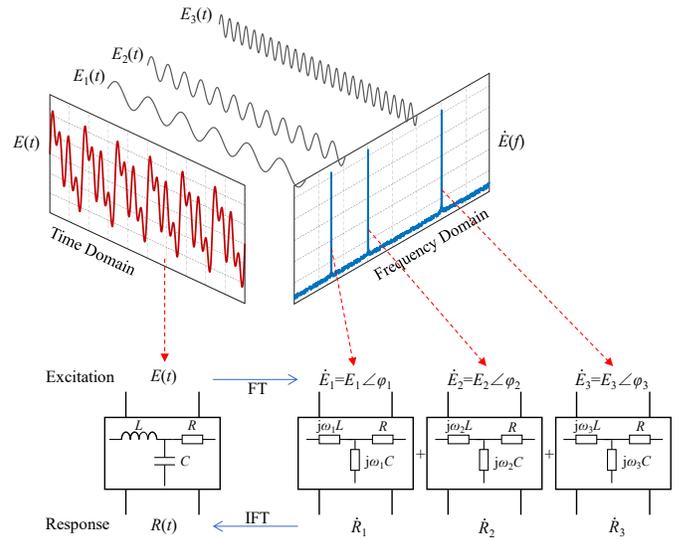

Fig. 3. The decomposition of excitations and the recovery of responses.

### Step 3) Two-Port Equivalence

Given the boundary values at the beginnings of pipelines, the FDDP equations of $\kappa$-th frequency component in (5) and (6) are analytically solved, whose solutions are as follows:

$$\begin{bmatrix} \dot{p}_\kappa(l) \\ \dot{m}_\kappa(l) \end{bmatrix} = \begin{bmatrix} \mathcal{A}_{f,\kappa} & \mathcal{B}_{f,\kappa} \\ \mathcal{C}_{f,\kappa} & \mathcal{D}_{f,\kappa} \end{bmatrix} \begin{bmatrix} \dot{p}_\kappa(0) \\ \dot{m}_\kappa(0) \end{bmatrix} \quad (7)$$

$$\begin{bmatrix} \dot{T}_\kappa(l) \\ \dot{h}_\kappa(l) \end{bmatrix} = \begin{bmatrix} \mathcal{A}_{t,\kappa} & \mathcal{B}_{t,\kappa} \\ \mathcal{C}_{t,\kappa} & \mathcal{D}_{t,\kappa} \end{bmatrix} \begin{bmatrix} \dot{T}_\kappa(0) \\ \dot{h}_\kappa(0) \end{bmatrix} \quad (8)$$

These two algebraic equations give the pi-equivalent FDLP circuits of natural gas pipelines and heating pipelines, as shown in the third column of Fig. 2. The expressions of transmission parameters $\mathcal{A}_{f,\kappa}$, $\mathcal{B}_{f,\kappa}$, $\mathcal{C}_{f,\kappa}$, $\mathcal{D}_{f,\kappa}$, $\mathcal{A}_{t,\kappa}$, $\mathcal{B}_{t,\kappa}$, $\mathcal{C}_{t,\kappa}$, and $\mathcal{D}_{t,\kappa}$, as well as that of equivalent lump branch parameters $\mathcal{Z}_{f,k}$, $\mathcal{Y}_{f1,\kappa}$, $\mathcal{Y}_{f2,\kappa}$, $\mathcal{K}_{f,\kappa}$, $\mathcal{Z}_{t,k}$, and $\mathcal{Y}_{t,\kappa}$, are provided in Appendix A.

networks, petroleum networks, steam networks, and hydrogen networks with minor modifications.

Regarding the mathematical derivations in the above three steps, we refer readers to our previous works [61]-[63] for more details.

*C. Energy Circuit Modeling for Networks*

As mentioned before, an energy network with non-sinusoidal excitations can be analyzed by converting original excitations into multiple sinusoidal components with different frequencies. Thus, we focus on the energy circuit modeling for networks under a single sinusoidal excitation component in this subsection, which is derived as follows. Note the subscript $\kappa$ that distinguishes different frequency components is omitted for the purpose of brevity.

1) **Branch Equation**

According to the pi-equivalent energy circuit models, each transmission line in electricity networks or pipeline in natural gas networks and heating networks can be modeled as three branches that are composed of impedance and controlled sources. Thus, a unified branch equation of these networks is given in matrix form as follows:

$$\dot{\mathcal{F}}_b = \mathcal{Y}_b (\dot{\mathcal{P}}_b - \mathcal{K}_b \dot{\mathcal{P}}_{bf}) \quad (9)$$

where, $\dot{\mathcal{F}}_b$ is a vector whose *i*-th entry is the flow variable of *i*-th branch, $\dot{\mathcal{P}}_b$ is a vector whose *i*-th entry is the potential difference of *i*-th branch, $\dot{\mathcal{P}}_{bf}$ is a vector whose *i*-th entry is the potential variable on the "from" side of *i*-th branch, $\mathcal{Y}_b$ is a diagonal matrix whose *i*-th entry is the admittance of *i*-th branch, and $\mathcal{K}_b$ is a diagonal matrix whose *i*-th entry is the pressure-controlled pressure source parameter of *i*-th branch. Particularly, $\mathcal{K}_b$ is a zero matrix for electricity networks and heating networks.

2) **Topology Equation**

The node-branch incidence matrix $\mathcal{A}$ and node-outflow-branch incidence matrix $\mathcal{A}_+$ are introduced to indicate the network topology, whose definitions are given below.

$$\mathcal{A}(i,j) = \begin{cases} 1, & \text{if branch } j \text{ flows from node } i \\ -1, & \text{if branch } j \text{ flows into node } i \\ 0, & \text{otherwise} \end{cases} \quad (10)$$

$$\mathcal{A}_+(i,j) = \begin{cases} 1, & \text{if branch } j \text{ flows from node } i \\ 0, & \text{otherwise} \end{cases} \quad (11)$$

where $\mathcal{A}(i,j)$ and $\mathcal{A}_+(i,j)$ are their entries at the *i*-th row and *j*-th column, respectively.

Based on these two matrices, the KCL and KVL constraints in electricity networks are generalized to the flow constraint and potential constraint. The flow constraint means that the difference between the outflow and inflow of a node equals its flow injection, as formulated in (12). The potential constraint correlates the branch potential and node potential, as formulated in (13) and (14).

$$\mathcal{A}\dot{\mathcal{F}}_b = \dot{\mathcal{F}}_n \quad (12)$$

$$\mathcal{A}^T \dot{\mathcal{P}}_n = \dot{\mathcal{P}}_b \quad (13)$$

$$\mathcal{A}_+^T \dot{\mathcal{P}}_n = \dot{\mathcal{P}}_{bf} \quad (14)$$

where $\dot{\mathcal{F}}_n$ is a vector whose *i*-th entry is the flow injection of *i*-th node and $\dot{\mathcal{P}}_n$ is a vector whose *i*-th entry is the potential variable of *i*-th node.

3) **Network Equation**

Substituting the topology equations (12)-(14) into the branch equation (9) yields a unified network equation for electricity networks, natural gas networks, and heating networks, as in (15).

$$\dot{\mathcal{F}}_n = \mathcal{Y}_n \dot{\mathcal{P}}_n \quad (15)$$

where $\mathcal{Y}_n = \mathcal{A}\mathcal{Y}_b\mathcal{A}^T - \mathcal{A}\mathcal{Y}_b\mathcal{K}_b\mathcal{A}_+^T$ is named the generalized node admittance matrix.

Thus far, the intractable PDEs that model natural gas networks and heating networks under general excitations have been equivalently transformed into a series of algebraic equations in the form of (15). Because this algebraization method is generalized from the electric circuit method, we name it the energy circuit method (ECM). Compared with the FDM that achieves the same function, the proposed ECM produces fewer variables and equations, which thereby yields better computation efficiency. This performance improvement is mainly attributed to avoiding spatial discretization by two-port equivalence in the ECM. In addition, different frequency components in the ECM are independent and therefore can be calculated in a parallel manner, while different time steps in the FDM have to be calculated serially. This distinction further widens the performance gap between the two methods.

It is worth noting that the ECM determines a general solution rather than a particular solution of the original PDEs if initial conditions are not specified. Considering the difficulty of specifying initial conditions in a frequency-domain model, an approach using historical boundary conditions as surrogate initial conditions so that the ECM can determine the desired particular solution is discussed in Appendix B.

III. IMPLEMENTATION: DESIGN AND DEVELOPMENT OF IEMS

Regarding the organization of this section, the design of the software architecture and function architecture for IEMSs is first introduced in subsections A and B, respectively. Then, the development of four typical advanced applications in IEMSs, including energy flow analysis, optimal energy flow, dynamic state estimation, and security assessment and control, is successively presented in the following four subsections of C, D, E, and F.

*A. Software Architecture Design*

Toward standardization and modularization, the hierarchical software architecture of IEMSs is designed as in Fig. 4. From a bottom-up perspective, this architecture contains five levels of operating system (OS), data, middleware, application, and front-end. A brief introduction to these levels is as follows.

1) **OS Level**

The IEMS is developed based on the off-the-shelf OSs. As a cross-platform software, it is compatible with Linux and Windows OS, which are most commonly used in industries.

2) **Data Level**

Diversified storage types, including real-time database (RTDB), SQL database, NoSQL database, and file storage, are integrated at the data level for different storage requirements. The RTDB responds to fast data-access requests during computation processes but has relatively small capacity, while the SQL and NoSQL databases undertake the function of large-capacity and persistent data storage, e.g., historical data and model data, but have relatively slow access speed. File storage is used for the purposes of program logs, data export, *etc*.

3) **Middleware Level**

At the middleware level, three buses of data, service, and message and eight modules of log management, warning management, process management, application management, resource monitoring, task dispatch, identity authentication, and security defence are developed to connect the low-level OS and data with high-level applications. The common services provided by these buses and modules greatly simplify the implementation and extension of advanced applications.

4) **Application Level**

Advanced applications, including SCADA, dynamic state estimation, generalized load prediction, energy flow analysis, security assessment and control, and optimal energy flow, are developed at this level, which provides the IEMS with functions covering the whole process of managing multiple energy flows. According to user needs, more applications, such as nodal price calculation and virtual power plants, can be extended.

5) **Front-end Level**

The front-end level provides a user interface that receives user commands to invoke the corresponding applications and displays the returned results. This user interface is implemented in three manners of local monitors, portable terminals (through a local area network), and remote web access (through a wide area network) to meet the diversified requirements of different scenarios.

*B. Function Architecture Design*

The function architecture of IEMSs, which organizes different advanced applications, is designed as in Fig. 5. The related details are as follows.

1) **SCADA**

SCADA collects real-time measurement data from sensors and performs data management, such as format regulation and indexed storage, to meet the data requirements of dynamic state estimation. Its detailed functions include data acquisition and processing, network topology coloring, event and alarm processing, automatic recording and printing, event recall and sequence recording, device remote control, *etc*.

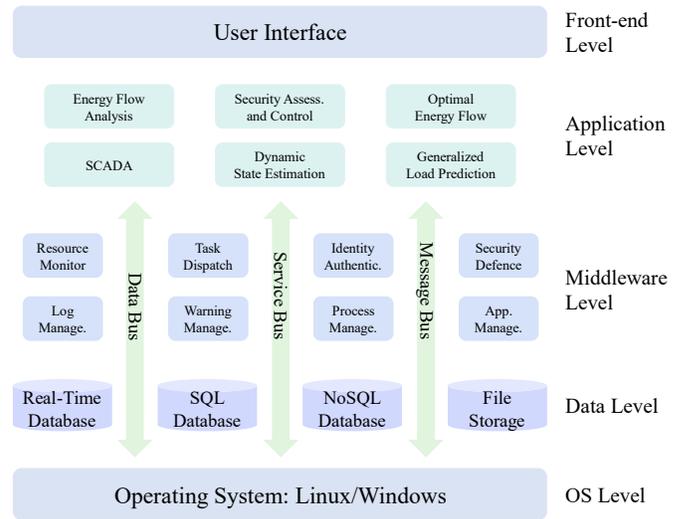

Fig. 4. Software Architecture of IEMSs.

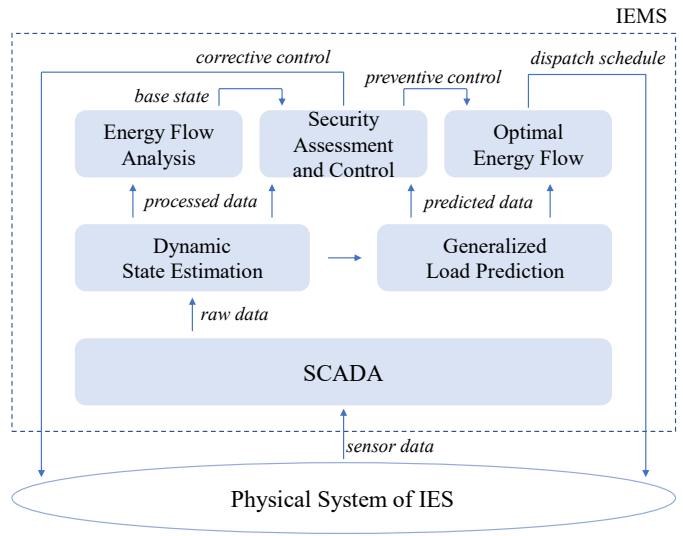

Fig. 5. Function architecture of IEMSs.

2) **Dynamic State Estimation**

Based on measurement data from SCADA, dynamic state estimation provides consistent, complete, and reliable estimated data of the historical states and current state, which satisfy both temporal and spatial constraints, for other applications. Its detailed functions include measurement noise filtering, bad data identification, pseudo measurement generation, observability analysis, network topology analysis, parameter recognition, *etc*.

3) **Generalized Load Prediction**

Using historical data and other information, such as weather and forecast day type, generalized load prediction forecasts multienergy loads and available renewable energy supplies in the future to provide important references for other applications formulating device schedules and assessing security risk. The predictions are categorized into time scales of middle-term, short-term, and super short-term according to different needs.

4) **Energy Flow Analysis**

With historical boundary conditions or current initial conditions from dynamic state estimation and future boundary conditions from generalized load prediction, energy flow analysis calculates temporal-spatial energy flow distributions in an IES from the present to a certain future. This application helps dispatchers perceive system trends in advance and provides a base state for further security assessment.

5) **Security Assessment and Control**

The application of security assessment and control warns and prevents security incidents within and across energy sectors. It generates a contingency set that comprises all failures of critical devices and then evaluates consequences of each contingency by invoking energy flow analysis to detect potential security constraint violations. If the assessment reveals a security risk, either preventive control that pulls the IES back into a normal and secure state before the contingency or corrective control that maintains the IES security after the contingency is adopted.

6) **Optimal Energy Flow**

Given forecasted multienergy loads and available renewable energy, optimal energy flow formulates dispatch schedules for all controllable devices, in which multienergy complementation and system inertia of natural gas networks and heating networks are totally considered to optimize the objective of operating costs, renewable energy accommodation, or energy supply efficiency. According to different calling periods and scheduling times, this application includes day-ahead, intra-day, and real-time scheduling models.

In these advanced applications, SCADA and generalized load prediction are uncorrelated to energy network modeling, so they can be developed in a manner similar to that in traditional EMSs, as in [64] and [65]. However, the remaining applications must be implemented considering the long-time-scale dynamic characteristics of natural gas networks and heating networks, which results in a significant difference from that in traditional EMSs. Their ECM-based implementations are introduced as follows.

*C. ECM-based Energy Flow Analysis Development*

1) **Related Works**

As the coupling between electricity, natural gas, and heating networks deepens and complicates, there is a trend to calculate energy flow distributions of different energy networks in a joint manner [35]-[37]. These works combine steady-state equations of various energy networks and derive a joint Jacobian matrix so that they can be solved by Newton-Raphson iterations. In [66], the computational performance is improved by replacing the original Jacobian matrix with a diagonal and constant one, which is inspired by the fast decoupled method for power flow calculation of electricity networks. Furthermore, a fixed-point iteration method and a holomorphic embedding method are employed to decompose the joint model for privacy reservation and calculation acceleration in [36] and [67], respectively. Considering uncertainties in generalized loads, probabilistic integrated energy flows are analyzed using the Monte-Carlo simulation, interval calculation method, and point estimation method in [68], [69], and [70], respectively. However, the above works all formulate natural gas networks and heating networks with steady-state models, which ignore the dynamic process from one steady state to another steady state, resulting in calculation errors. In fact, large-scale natural gas networks and heating networks do not even enter a steady state under hour-level or minute-level regulation. For this issue, integrated energy flow calculations with dynamic models embedded are studied in [26] and [71]-[72]. However, the PDEs of dynamic models are all algebraized by the FDM in these works, which consequently brings a considerable computational burden.

2) **ECM-based Energy Flow Analysis**

In essence, the ECM-based energy flow analysis solves the following simultaneous equations: (1) time-domain power flow equations of electricity networks, (2) frequency-domain fluid circuit equations of natural gas networks, (3) frequency-domain thermal circuit equations of heating networks, (4) time-domain energy conversion equations, and (5) time-frequency transformation equations. Their mathematical formulation is provided in Appendix C.

Since the above five equations are all algebraic equations, the Newton-Raphson method is a fundamental option to find the solution. However, a huge Jacobian matrix containing variables from three energy networks and two domains imposes a heavy computational burden on the matrix inversion operation. A better approach is using the fixed-point method to iterate boundary variables so that the original problem is decomposed into three subproblems where each subproblem contains only variables of one energy network. For the electricity network subproblem, power flow equations of different time steps are solved in a parallel manner. For the natural gas network subproblem and the heating network subproblem, the time-domain boundary conditions are transformed into the frequency domain and thereby fluid circuit equations and thermal circuit equations are solved in a parallel manner. More details about solving fluid circuit equations and thermal circuit equations can be found in our previous work [73].

3) **Case Study**

Two test systems are prepared for validation. The first one is a small-scale IES for illustration purposes, which contains a 9-bus electricity network, a 7-node natural gas network, and a 12-node heating network. The topology of networks and devices, including a wind turbine, a combined heat and power (CHP) unit, a gas turbine, two gas wells, and a gas boiler, is shown in Fig. 6. The second one is a large-scale IES for performance verification purposes, which contains a 118-bus electricity network, a 181-node natural gas network, and a 376-node heating network. The natural gas network and heating network in this test system are modified from two actual city-scale energy networks.

The energy flow distributions of the illustrative test system in 24 hours with an interval of 5 minutes are calculated. The results of transmission line power, pipeline temperature, node pressure, and pipeline gas flow are shown in Fig. 7, in which the solid lines are computed by the ECM using historical boundary conditions of 24 hours as surrogate initial conditions and the dotted lines are computed by the FDM with the implicit upwind scheme and parameters used in [40]-[41]. Regarding the results of the FDM as benchmarks, the average maximum error of the ECM is within 0.62%. With such accuracy, the elapsed time of the ECM is only 0.34 s, which is far less than the 14.75 s of the FDM.

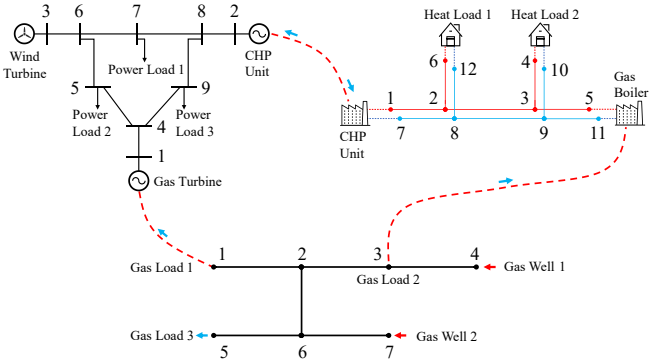

Fig. 6. Topology of the illustrative test system.

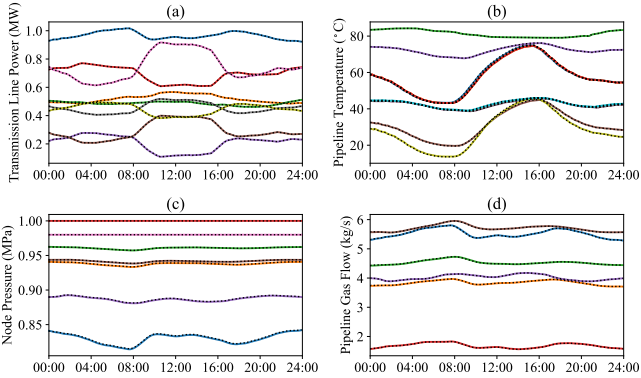

Fig. 7. Energy flow distributions of the illustrative test system: (a) power on transmission lines, (b) temperature at the end of heating pipelines, (c) pressure of nodes in the natural gas network, and (d) gas flow in natural gas pipelines. The solid lines are computed by the ECM, and the dotted lines are computed by the FDM (they are almost overlapped).

The two methods are further compared on the large-scale test system to demonstrate the computational advantage of the ECM. For 24-hour energy flow analysis, under nearly the same accuracy, the ECM costs only 1.30 s, while the FDM costs 251.51 s. This result indicates that the ECM-based IEMS is able to perform online energy flow analysis on city-scale IESs.

### D. ECM-based Optimal Energy Flow Development

#### 1) Related Works

As pioneering works on optimal energy flow, electricity networks and natural gas networks are coordinated to eliminate gas supply shortages in [74], while electricity networks and heating networks are coordinated to mitigate wind curtailment in [75]. A comprehensive model that simultaneously optimizes power, gas, and heat is formulated in [76]. These works are implemented using the steady-state models of natural gas networks and heating networks, which are replaced by the dynamic models to utilize the inertia of these two networks as energy storage in [77] and [78]. Interpreted from mathematics, the latter works search for a better solution in a larger feasible region. Currently, the PDE constraints in dynamic optimal energy flow models are mostly algebraized by the FDM, which significantly scales up the size of optimization problems.

More complex factors are considered and addressed in later studies. Considering the privacy issue, the joint optimization is solved in a decentralized manner using the multiagent genetic algorithm [79], generalized Bender's decomposition [80], and alternating direction method of multipliers [81], which ensures that the parameters of different energy networks are unshared. Considering individual incentives, a Nash-Stackelberg game model [82] and a transfer payment strategy [83] are proposed for the optimal energy flow problem in which different energy networks are managed by independent entities. Considering uncertainties, stochastic optimization [84], robust optimization [85], and distributionally robust optimization [86] are employed to give a deterministic dispatch schedule. The proposed energy circuit model is compatible with these works.

#### 2) ECM-based Optimal Energy Flow

The ECM-based optimal energy flow is an optimization problem, in which decision variables are divided into four categories: time-domain controlled variables, time-domain monitored variables, frequency-domain controlled variables, and frequency-domain monitored variables. The time-domain controlled variables are the outputs of controllable devices, such as generator power and gas well production, which formulate the objective function of minimizing the total cost and satisfy the operation constraints of devices including output ranges, ramping limits, and energy conversion relationships. The time-domain monitored variables are the network states determined by device outputs, such as transmission line power, node pressure, and pipeline temperature, which satisfy the security constraints of networks. Instead of using steady-state network constraints or FDM-based dynamic network constraints to couple time-domain controlled variables and time-domain monitored variables as in conventional works, ECM-based dynamic network constraints are first employed to couple controlled variables and monitored variables in the frequency domain, and then time-frequency transformation constraints are formulated to couple time-domain variables and frequency-domain variables. This distinction is illustrated as in Fig. 8.

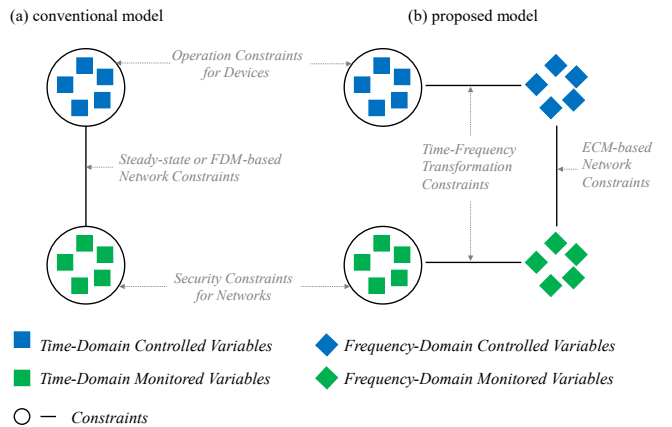

Fig. 8. Comparison between the conventional and proposed optimal energy flow models.

The final optimization model, whose complete mathematical formulation is provided in Appendix D, is a quadratic programming problem and can be efficiently solved using off-the-shelf commercial solvers such as Gurobi [87] and CPLEX [88]. To further compact the optimization model and accelerate solving, more techniques, including variable space projection and row generation, have been developed. We refer readers to our other work [63] for more details about these techniques.

3) **Case Study**

The day-ahead optimal energy flow with an interval of 15 minutes is solved for the illustrative test system in Fig. 6 using the ECM-based dynamic model, FDM-based dynamic model, and steady-state model, whose results are reported as follows.

Regarding the comparison between the dynamic and steady-state models, the related results are summarized in TABLE II, and dispatch schedules of three energy networks are provided in Fig. 9. For the steady-state gas network model, there is no feasible solution since the two gas wells cannot cover the peak gas load at approximately 10:00. In contrast, the dynamic gas network model gives a feasible dispatch schedule that allows mild mismatch between the gas load and gas supply at the cost of acceptable pressure fluctuation. For the steady-state heating network model, the heat supply follows the heat load plus the heat loss. Consequently, the CHP unit maintains high output during the night (00:00-04:00) due to heavy heat load, which squeezes the space for wind power accommodation through heat-power coupling. In addition, the low output of the CHP unit during the daytime (10:00-14:00) results in more power supplied by the gas turbine, which is more expensive than the CHP unit. In contrast, the dynamic heating network model avoids the above two circumstances by shifting the peak load to the valley load and finally gives a dispatch schedule with less wind power curtailment and operating cost.

The FDM-based dynamic model gives a dispatch schedule similar to the ECM-based dynamic model, and the difference of their optimal objectives is within 0.1%. With almost the same optimality, the ECM obviously outperforms the FDM on efficiency: the ECM-based model contains nearly 95% fewer variables and constraints and is solved in roughly 75% less time than the FDM-based model. Details are listed in TABLE III.

In the performance validation on the large-scale test system, the ECM costs only 134 s to solve the day-ahead optimal energy flow, which is far more efficient than the FDM that costs 1615 s. This result confirms that the ECM-based IEMS is able to perform online energy flow optimization on city-scale IESs.

TABLE II
RESULT COMPARISON OF DYNAMIC AND STEADY MODELS

|  | Dynamic Model | Steady Heat Model | Steady Gas Model |
|---|---|---|---|
| Operating Cost ($10^3$ \$) | 346.60 | 357.09 | Infeasible |
| Wind Power Curtail. (%) | 5.47 | 17.71 |  |
| Heat Loss (%) | 6.42 | 7.70 |  |

TABLE III
PERFORMANCE COMPARISON OF ECM AND FDM

|  |  | ECM | FDM |
|---|---|---|---|
| Problem Scale | Num. Variables | 10,040 | 212,448 |
|  | Num. Constraints | 14,832 | 232,621 |
| Elapsed Time (s) | Modeling Time | 4.10 | 15.73 |
|  | Solving Time | 4.30 | 19.48 |
| Optimal Objective ($10^3$ \$) |  | 346.60 | 346.33 |

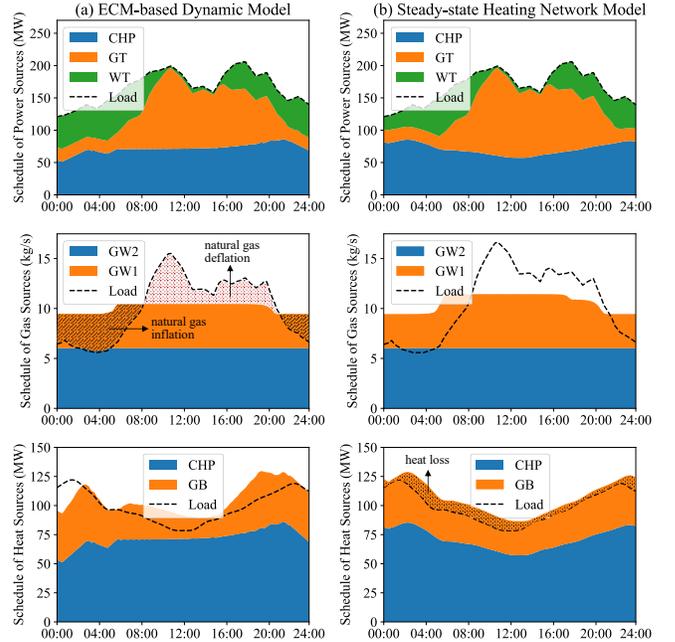

Fig. 9. Optimal schedules computed by (a) the ECM-based dynamic model and (b) the steady-state heating network model.

*E. ECM-based Dynamic State Estimation Development*

1) **Related Works**

State estimation that improves measurement accuracy has been broadly studied and applied in power systems [89]. With increasingly tight coupling of different energy networks, related works have been expanded to IESs for joint state estimation of power, gas, and heat [90]-[93]. The state estimation models in these works are established based on the weighted least square method. Considering more complex factors in IESs, a weighted least absolute value-based state estimation model is formulated in [94], and another second-order cone programming-based state estimation model is investigated in [95]. Furthermore, a decentralized state estimation method for IESs is proposed in [96] to address the privacy issue.

However, the above works have followed the section-based estimation manner adopted in electricity networks, i.e., using steady-state network models and current measurement data to estimate the current state, which ignores dynamic processes and results in obvious estimation error. To close this research gap, dynamic state estimation models, in which both historical and current measurement data are utilized to estimate the current state, are developed for heating networks and natural gas networks in [97]-[98] and [99]-[100], respectively. As a common practice, the PDE constraints on dynamic processes are algebraized by the node method and FDM in extant works on dynamic state estimation.

2) **ECM-based Dynamic State Estimation**

The ECM-based dynamic state estimation is an optimization problem, which seeks both historical and current states with minimum deviation from measurement data under the premise of satisfying temporal-spatial physical equations for energy networks. Note that the estimation for historical states is a customized requirement of the ECM-based IEMS, since the ECM needs historical boundary conditions to surrogate initial conditions.

Specifically, the decision variables include time-domain state variables and frequency-domain state variables, in which the former are main variables to be solved and the latter are auxiliary variables to introduce energy circuit models for efficient algebraization. The objective function is usually minimizing the weighted sum of measurement residuals. The constraints include the AC power flow model of electricity networks, fluid circuit model of natural gas networks, thermal circuit model of heating networks, time-frequency transformation constraints, and energy coupling constraints between different energy networks. Their mathematical formulation is provided in Appendix E.

The final optimization model involves nonlinear equality constraints from the AC power flow model. Thus, the nonlinear programming solver IPOPT [101] that is based on the primal-dual interior point method is adopted for solving. Although the above joint state estimation can guarantee a global consistent state across different energy networks in an IES, sequential state estimation, i.e., estimate states of electricity networks first and then estimate states of natural gas networks and heating networks in which the estimation results of electricity networks are regarded as pseudo measurements of the other two networks through energy coupling relationships, is also recommended due to the following two considerations:

i) Electricity networks require more frequent state estimation than natural gas networks and heating networks because of their faster dynamic processes.

ii) Sequential state estimation separates nonlinear constraints of electricity networks and linear constraints of natural gas networks and heating networks, which contributes to improving the solving efficiency.

3) **Case Study**

The steady state estimation (SSE), FDM-based dynamic state estimation (FDM-DSE), and ECM-based dynamic state estimation (ECM-DSE) are performed on the illustrative test system in Fig. 6. The test configuration is as follows: the estimation time is 24 hours, the measurement interval is 5 minutes, the true values of states are calculated by energy flow analysis, and the measurement data are randomly generated by superposing Gaussian noise with an amplitude of 0.01 p.u. on the true values. Using the ratio of squared error of estimation values to squared error of measurement values (denoted as $\xi$) as accuracy metrics, the three methods are compared in TABLE IV. It is observed that the average value of $\xi$ obtained by the SSE is larger than 1 for both the natural gas network and heating network, which indicates that the estimation data of the SSE are even worse than the measurement data. In contrast, the average values of $\xi$ obtained by the FDM-based DSE and ECM-based DSE are both smaller than 1, which validates their effectiveness in filtering measurement noise. Particularly, the ECM-based DSE outperforms the FDM-based DSE on accuracy because the limited frequency components in the ECM implicitly filter part of high-frequency noise.

Some typical estimation results of the SSE and ECM-based DSE are provided in Fig. 10, which indicates that the curves estimated by the SSE fail to track the dynamic processes and have many glitches (i.e., unsmooth) due to less consideration of the temporal couplings. These phenomena explain why the SSE produces a large $\xi$.

TABLE IV
ACCURACY COMPARISON OF THREE ESTIMATION METHODS

|  | SSE | | FDM-DSE | | ECM-DSE | |
| --- | --- | --- | --- | --- | --- | --- |
|  | HN | NGN | HN | NGN | HN | NGN |
| Avg. $\xi$ | 1.67 | 2.07 | 0.22 | 0.41 | 0.02 | 0.07 |
| Max. $\xi$ | 8.91 | 10.21 | 0.63 | 0.92 | 0.21 | 0.66 |

* HN: heating network, NGN: natural gas network

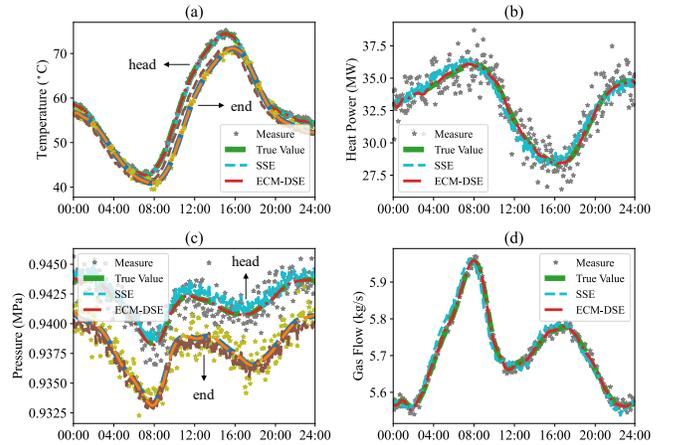

Fig. 10. Estimation results of the SSE and ECM-based DSE: (a) temperature at the head and end of heating pipeline 1-2, (b) heat power of the gas boiler, (c) pressure at the head and end of gas pipeline 3-4, and (d) average gas flow of gas pipeline 1-2.

In addition to better estimation accuracy, the ECM-based DSE spends 0.87 s on calculation, which is obviously less than the 4.38 s of the FDM-based DSE.

*F. ECM-based Security Assessment and Control Development*

1) **Related Works**

Security assessment that evaluates system robustness to disturbance has been widely studied for power systems, as reviewed in [102]. Since the coupling between different energy flows may bring cascading failures across energy sectors, security assessment for IESs has been explored, such as works in [103] and [104], which calculate integrated energy flows after each failure in the contingency set to confirm whether an

IES is secure. For faster online use, [105] and [106] propose an IES security region method, but it is rather conservative for practical use. These works focus on the post-contingency steady state but ignore the transient process between pre-contingency and post-contingency steady states. To provide this missing information, it is necessary to carry out dynamic energy flow analysis in security assessment for IESs.

If any security violation is detected, security control regulates controllable devices to maintain system security, which is mainly divided into preventive control and corrective control according to when the control is performed. The former modifies device schedules before failure to ensure that the post-contingency state is still secure, whereas the latter modifies device outputs after failure to prevent system collapse. These security control problems for power systems are investigated in [107], and similar works on preventive control [74], [108] and corrective control [109]-[110] have been developed for IESs. Nevertheless, the research gap of considering dynamic characteristics in the security control for IESs still remains.

2) **ECM-based Security Assessment and Control**

The ECM-based security assessment contains two steps of contingency set generation and post-contingency energy flow analysis. Regarding the first step, its implementation is similar to that in conventional EMSs [111]. Regarding the second step, the aforementioned ECM-based energy flow analysis is utilized to obtain temporal-spatial dynamic energy flow distributions after a given failure so that both whether and when the IES is insecure after the failure can be determined.

The ECM can directly calculate post-contingency energy flow of device contingencies that influence network injections, such as the outages of generators, boilers, and gas wells, as long as the node injection corresponding to the failure device is set to zero after the fault time. However, it is rather intractable for the ECM to address branch contingencies that influence network topology, such as the outages of gas pipelines, because two topologies before and after the contingency yield a time-variant admittance matrix that invalidates the superposition theorem and thereby non-sinusoidal excitations cannot be decomposed for independent calculation on sinusoidal components. Fortunately, such a time-variant problem can be transformed into a time-invariant problem by replacing the branch removal with two equivalent node injections, as used when deriving line outage distribution factors for power systems [112]. Specifically, this strategy first calculates the pre-contingency energy flow. Then, two imaginary injections are added to the nodes connected by the outage branch: the injection at the inflow node before the fault time is zero, and that after the fault time is the flow on the outage branch *after* the imaginary injections are imposed, which equals the pre-contingency value multiplied by a scale factor[4]; the injection at the outflow node is the opposite of that at the inflow node. It is worth noting that this equivalent replacement, as visualized by an illustrative example in Fig. 11, is only adopted for natural gas networks and heating networks to ensure a time-invariant admittance matrix so that our superposition theorem-based ECM is applicable. For transmission line outages in electricity networks, the post-contingency energy flow is calculated using a new admittance matrix in which the outage branch is excluded.

Preventive control or corrective control is adopted according to the severity and rapidity of failure consequences. Their ECM-based implementations are similar to optimal energy flow except for the following differences. For preventive control, there is one more post-contingency security constraint, which is formulated by energy circuit-based sensitivity factors [113]. For corrective control, the objective function switches to minimizing state deviations so that the new secure state can be reached quickly, and there are some additional constraints to reflect the occurred failure.

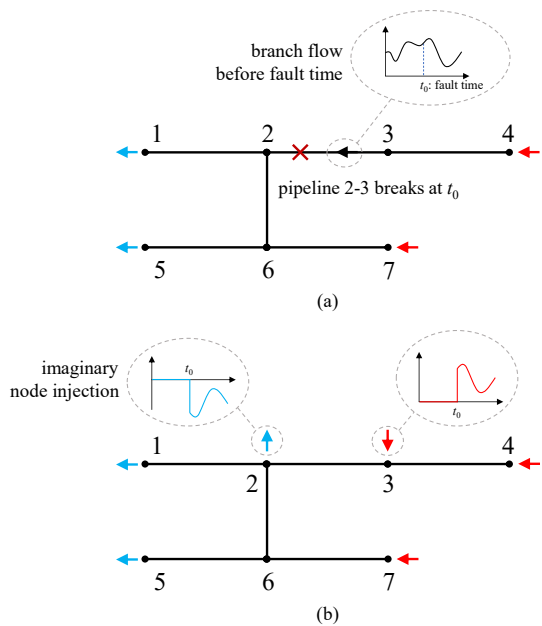

Fig. 11. Replacing a branch removal with two node injections: (a) original time-variant topology with a pipeline outage at $t_0$, (b) equivalent time-invariant topology with two imaginary node injections.

3) **Case Study**

The *N*-1 contingency set of the illustrative test system in Fig. 6 contains 26 failures, from which a device contingency of gas turbine outage and a branch contingency of gas pipeline 2-3 outage are selected to test the ECM-based security assessment. Setting the fault time at 12:00, the results are shown in Fig. 12.

Although the gas turbine outage does not result in overload or loss of load in the electricity network, load temperature exceeding the upper limit occurs in the heating network. The mechanism of fault propagation across energy sectors is as follows: the power mismatch induced by the gas turbine outage is all compensated by the CHP unit, which in turn increases the heat supply due to the coupling. Finally, the over-supplied heat leads to a global temperature rise in the heating network. After

---

[4] We refer readers to the Appendix 11A in [112] for more details about this factor and its derivation.

63 minutes, the nearest load 1 first touches the security limit of 120 °C. The details are given in Fig. 12 (a).

After the outage of gas pipeline 2-3, the gas supply path of load 1 is switched from "pipeline 3-4 > pipeline 2-3 > pipeline 1-2" to "pipeline 6-7 > pipeline 2-6 > pipeline 1-2", which significantly aggravates the gas flow on pipeline 6-7 that has already delivered the gas flow of load 3. As a consequence, the pressure loss on pipeline 6-7 increases and the pressure at load 1 and load 3 decreases. Due to large inertia of the natural gas network, this flow-shifting process lasts several hours, and the furthest load 1 first touches the security limit of 0.7 MPa after 202 minutes. The details are presented in Fig. 12 (b). Because gas load 1 supplies the gas turbine, this under-pressure fault further trips the gas turbine and leads to another incident in the heating network, i.e., the first failure introduced above.

The FDM-based security assessment gives close results: the security constraint violations occur 55 minutes after the first failure and 212 minutes after the second failure. However, its calculation time is nearly 15 times to the ECM-based security assessment.

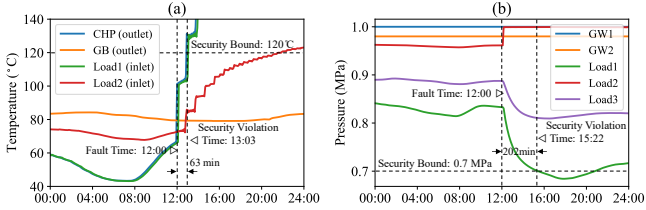

Fig. 12. Security assessment results: (a) temperature of heat sources and heat loads when the gas turbine trips at 12:00, (b) pressure of gas wells and gas loads when pipeline 2-3 in the natural gas network breaks at 12:00.

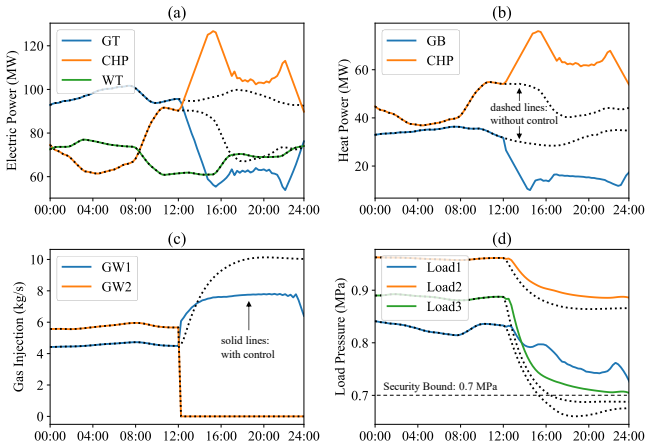

Fig. 13. Corrective control results for the outage of gas pipeline 3-4 at 12:00: (a) schedules of power sources, (b) schedules of heat sources, (c) schedules of gas sources, and (d) pressure of gas loads.

Another failure that causes inadequate gas supply pressure is the outage of gas pipeline 3-4, which is equivalent to the outage of gas well 1. The load pressure under the original schedules after this failure occurred at 12:00 is presented by the dotted lines in Fig. 13 (d), which indicate the pressure of load 1 and load 3 exceeds the lower limit at 15:16 and 16:28, respectively. To avoid such violation, corrective control is performed to maintain security, which gives the modified schedules in Fig. 13 (a), (b), and (c) by the solid lines. The corrective mechanism is as follows: raising the power and heat outputs of the CHP unit leads to the gas turbine and gas boiler decreasing their outputs, so as their gas consumptions, which thereby alleviates the gas transmission congestion in the natural gas network. As a consequence, the gas load pressure is maintained above the security lower bound, as shown in Fig. 13 (d).

## IV. APPLICATION: ENGINEERING DEMONSTRATIONS IN CHINA

Regarding the organization of this section, an overview of IEMS engineering demonstrations in China is first reported in subsection A, which follows detailed introductions to two typical demonstration projects for IESs with different scales in subsections B and C, respectively.

### A. Overview of IEMS Engineering Demonstrations

Since 2016, our developed IEMSs have been deployed onsite for engineering demonstrations. By the middle of 2022, a building-scale IES, 5 park-scale IESs, and 2 city-scale IESs have been equipped with IEMSs, with an additional 2 park-scale IESs and 2 city-scale IESs under construction for IEMSs. These buildings, commercial or industrial parks, and cities for IEMS engineering demonstrations are located in 10 provinces of China, which have different energy endowments and climatic conditions to provide various scenarios for verification of the IEMS. A partial list of the implemented and ongoing engineering demonstration projects is compiled in TABLE V. Furthermore, another building-scale IES, 5 park-scale IESs, and 6 city-scale IESs will deploy an IEMS, which are currently in the planning stage.

In these engineering demonstrations, the IEMS covers the functions of monitoring, analysis, optimization, and control on multiple energy flows of electricity, natural gas, steam (for industrial heating purposes), hot water (for residential heating purposes), and cold water (for residential cooling purposes). Through the coordinated operation of these energy flows, the IEMS has achieved significant economic and environmental benefits, which include not only the primary benefits such as reducing energy-consumption costs, promoting renewable energy accommodation, inhibiting peak loads, and improving energy-supply reliability, but also the secondary benefits such as reducing or delaying investment for energy-supply facilities and reducing carbon emissions. The obtained benefits in the implemented engineering demonstrations are quantitatively summarized in TABLE V.

Throughout the development of IEMSs for these engineering demonstrations, the gradual expansion of IES scales has brought about two changes to the network modeling in IEMSs. In the first period, the engineering demonstrations were carried out in buildings and commercial parks whose pipeline networks are so small that the regulation on the supply side immediately changes the state on the demand side. Therefore, steady-state network equations were chosen to model these small-scale

IESs. In the second period, our IEMS began to manage multiple energy flows in large industrial parks and towns, in which obvious dynamic processes inside pipeline networks were observed. To involve this feature in modeling, the FDM-based dynamic network equations were chosen for these middle-scale IESs and worked well. In the third period, however, the FDM yields numerous discrete elements when modeling city-scale IESs, which severely slows the analysis and decision-making process of IEMSs. In such a circumstance, the ECM was proposed to enable IEMSs to perform online management on large-scale IESs.

TABLE V
PARTIAL LIST OF IEMS ENGINEERING DEMONSTRATIONS IN CHINA

| Location | IES Scale | Status | Energy Sectors | More Available Details |
|---|---|---|---|---|
| Changchun, Jilin | city | implemented | electricity heating | A heating network with over 268 km of pipelines that supplies heat for a total area of 20 million $m^2$ and an electricity network with 97 220-kV substations and wind turbine installed capacity of 5,050 MW are involved. After the IEMS is accessed, the wind power curtailment in the demonstration area decreases by 12.65%, and the daily operating cost reduces by 4.87%. |
| Zhejiang | city | under construction | electricity natural gas | A natural gas network with over 1,600 km of pipelines that supplies natural gas for 11 cities of Zhejiang Province is involved. |
| Dehong, Yunnan | city | under construction | electricity natural gas steam | A cross-border IES, which includes a city-level electricity network, a city-level natural gas network, and several industrial parks with multiple energy loads in China and a city-level electricity network in Myanmar, is involved. |
| Beichen, Tianjin | park | implemented | electricity natural gas heating/cooling | A town that covers an area of 20.3 $km^2$ and has abundant flexibility resources, including CHP units, gas boilers, ground-source heat pumps, and energy storage devices, is involved. After the IEMS is accessed, the daily operating cost of this town reduces by 6.3% and the comprehensive energy efficiency increases by 5.76%. |
| Yangzhong, Jiangsu | park | implemented | electricity natural gas | A town that covers an area of 49.2 $km^2$ and has an installed renewable energy capacity of 150 MW is involved. After the IEMS is accessed, the comprehensive energy efficiency of this town increases by 3.03% and the local renewable energy accommodation reaches 100%. |
| Conghua, Guangzhou | park | implemented | electricity steam cooling | An industrial park that covers an area of 12 $km^2$ and has a power load of 20 MW, a steam load of 50 t/h, and a cooling load of 15 MW is involved. After the IEMS is accessed, the peak power load reduces by 20% through introducing demand response, which saves roughly 48 million yuan of investment on energy-supply facilities, and the carbon emission reduces by 6,000 tons per year. |
| Beike, Beijing | park | implemented | electricity natural gas heating/cooling | An industrial park that covers an area of 70 acres and has a power load of 4 MW and a heating/cooling load of 5 MW is involved. After the IEMS is accessed, the daily comprehensive energy consumption of this park reduces by 3%, which contributes to the operating cost saving of 512 thousand yuan per year. |
| Huaneng, Jilin | building | implemented | electricity heating/cooling | A commercial building that has a total area of 16,000 $m^2$ and a heating/cooling load of 1.5 MW is involved. After the IEMS is accessed, the daily operating cost has been reduced by 4.3%. |

*B. An Engineering Demonstration in Beike Industrial Park*

A typical demonstration project of the IEMS in Beike industrial park, which involves a park-scale IES, is reported as follows.

1) **Project Background**

Beike industrial park is located in the Haidian district, Beijing, which covers an area of 70 acres and has 6 buildings with a total construction area of 63,152 m$^2$. Fourteen companies, involving industries of opto-mechatronics, new material manufacturing, bioengineering and medicine production, *etc.*, have settled in this park and formed increasing demands for electricity, natural gas, heating, and cooling. By 2017, the peak loads have reached 4,000 kW for electricity and 5,000 kW for heating/cooling, which approach the designed capacity of the original park energy system. In order to better satisfy the energy demands of companies in the park, the park energy system was renovated in 2017 by introducing the facilities including a combined cold, heat, and power (CCHP) unit, a direct gas-fired heater and chiller, two gas boilers, three rooftop photovoltaic (PV) systems, a micro-wind turbine, and two battery storage systems. A location map of Beike industrial park and these facilities is provided in Fig. 14.

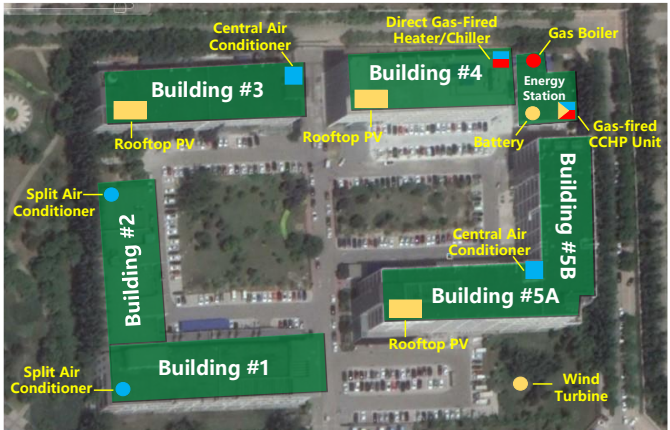

Fig. 14. A location map of Beike industrial park and related facilities.

On the basis of improvements on hardware facilities, a demonstration project that aims at developing an IEMS for such a park-scale IES to realize refined management on integrated energy flows is conducted.

2) **Field-Site Deployment**

The energy topology of the Beike industrial park is shown in Fig. 15, which presents a deep coupling relationship between different energy flows. As indicated, the sensor data of the related facilities, pipelines, and terminal users are collected by the IEMS so as to perform joint analysis, optimization, and control. Specifically, panoramic perception of the whole park energy system is provided through SCADA and state estimation, whose typical user interface is shown in Fig. 16; the operation schedules of facilities are formulated by the optimal energy flow application to achieve the most economical energy-supply scheme; potential security risks are recognized through security assessment, which helps park operators find vulnerable spots of the system and prepare emergency plans in advance; and the aggregation of distributed flexibility resources is supported through the virtual power plant application, which enables the park to respond to the peak shaving demand of the upper utility grid for extra profits. Because of the small scale of this park, the above functions are all executed in seconds.

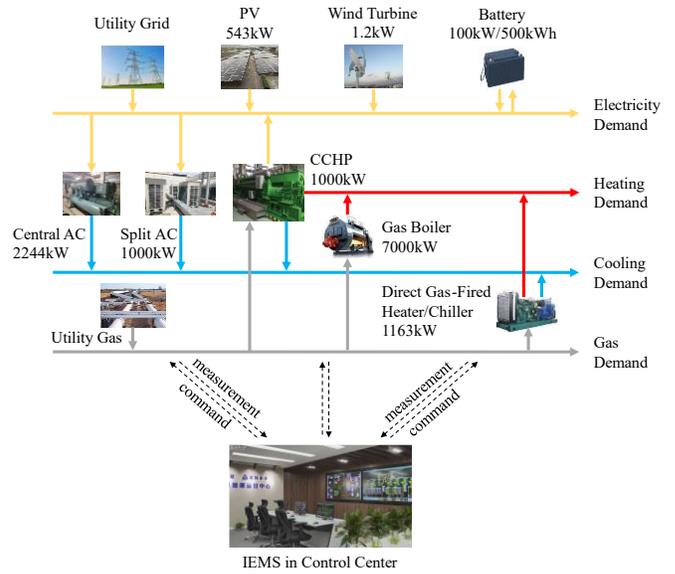

Fig. 15. Deployment relationships of the IEMS and other facilities in the Beike demonstration project.

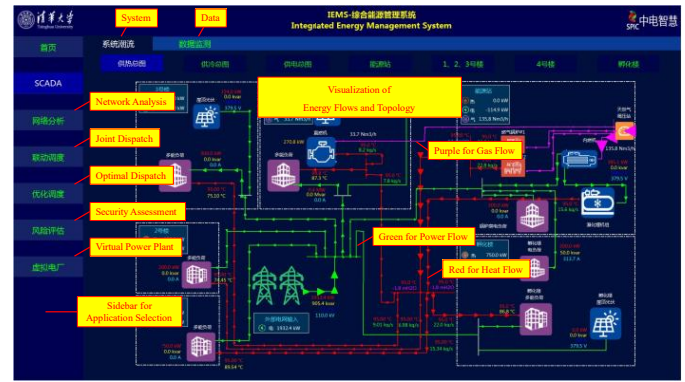

Fig. 16. Typical user interface of the park-scale IEMS.

3) **Overall Benefits**

The benefits of this demonstration project are summarized in the following three points.

First, the IEMS realizes the automation of energy supply, which relieves the work intensity of park operators. For instance, the visualization of energy flows and monitoring logs bring great convenience for operators troubleshooting failure causes.

Second, the IEMS coordinates different energy flows to decrease the total operating cost. After Beike industrial park is accessed to the IEMS, its daily comprehensive energy consumption has been reduced by 3%, which is equivalent to an annual saving of 512 thousand yuan per year.

Finally, the IEMS enables the park to participate in the ancillary service market of north China as a virtual power plant, which brings the park extra profits and helps the utility grid accommodate more renewable energy.

Another similar park-scale IEMS is deployed in Conghua industrial park, Guangzhou, whose introduction can be found in our previous work [1].

*C. An Engineering Demonstration in Jilin Province*

A typical demonstration project of the IEMS in Jilin Province, which involves a city-scale IES, is reported as follows.

1) **Project Background**

Jilin Province is located in the northeast of China. By the end of 2017, the electricity network in this province included 97 220-kV substations, 12,965 km transmission lines, and a total installed capacity of 26,860 MW. To utilize the rich wind resources in this province, wind turbines with a capacity of 5,050 MW are installed, which equals 18.80% of the total installed capacity, 50% of the maximum load, and 123% of the minimum load. In 2017, the power generated by wind turbines reached 8.69 billion kWh. Due to its high latitude, this province has a 6-month heating season from October to April of the following year, and multiple district heating networks with pipelines of a total length over 28,000 km have been built to supply heat for a total area of 611.5 million $m^2$. More than 70% of the heat load is supplied by CHP units operating with a heat-led strategy, which significantly limits the peak-shaving ability of these CHP units and squeezes the space for wind power accommodation. As a result, the annual wind power curtailment rates in Jilin Province were always above 20% and even 30% in 2017 and before, in which the curtailment during heating seasons accounted for 80%-90%.

In response to this situation, a demonstration project supported by the Ministry of Science and Technology of China and the State Grid Corporation of China is carried out. This project contains two phases: the first phase mainly focuses on reducing wind power curtailment by coordinated dispatch of the electricity network and heating network (i.e., the optimal energy flow application of the IEMS), while the second phase mainly focuses on the comprehensive energy management of the two energy networks (i.e., the remaining applications of the IEMS). This demonstration project aims to improve both wind power accommodation in the electricity network and the operation management level of the heating network.

2) **Field-Site Deployment**

The engineering demonstration is implemented in an area of Changchun, which is the capital of Jilin Province. In this demonstration area, 8 CHP units with an installed capacity of 1,820 MW, 8 wind farms with an installed capacity of 1,288 MW, and 2 electric-heating substations with an installed capacity of 46 MW are integrated to supply power and heat. The heat is delivered by a heating network that has 400 heat-exchange stations and pipelines over 268 km (containing water of roughly 300 thousand tons) to a load of 20 million $m^2$.

The deployment relationships of the IEMS and these facilities are illustrated in Fig. 17: the sensor data of the electricity network and power-supply facilities are collected by the data server in the electric power control center (EPCC), while those of the heating network and heat-supply facilities are collected by the data server in the heat power control center (HPCC). These measurement data, as well as other necessary data such as topology and contingencies, are sent to the IEMS server that is located in the EPCC for calculations. The results are sent back to the data servers in the EPCC and HPCC, some of which are issued to facilities for adjusting their outputs and the other of which are stored for queries from the work stations in the control centers.

The power and heat system involved in this demonstration project contains a 621-bus-817-branch electricity network and a 376-node-508-branch heating network. Using the ECM-based implementation, the calculation performance of advanced applications in the IEMS all meets the online-use requirements. Specifically, the dynamic state estimation is executed every 5 minutes, and each execution is finished in 30 seconds (typical); the energy flow analysis is executed every 5 minutes, and each execution is finished in 5 seconds (typical); the security check is executed every 60 minutes, and each execution is finished in 20 minutes (typical); the day-ahead optimal energy flow is executed once a day, and each execution is finished in 3 minutes (typical); and the intraday-rolling optimal energy flow is executed every 4 hours, and each execution is finished in 3 minutes (typical).

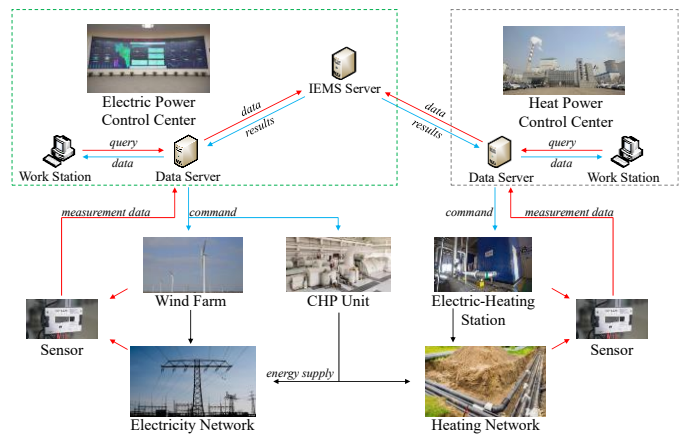

Fig. 17. Deployment relationships of the IEMS and other facilities in the Jilin demonstration project.

3) **Overall Benefits**

The heat-storage characteristics of massive water in the heating network provide significant flexibility for wind power accommodation. A typical scenario is provided in Fig. 18, which is a screenshot of the developed IEMS on March 1st, 2018. On this day, there was a large output of wind power from 00:00 to 05:00. To accommodate excessive wind power, the outputs of CHP units were decreased, which led to a 5 ℃ drop in the water supply temperature. However, due to the large inertia of the heating network, little influence on the indoor temperature caused by this adjustment was observed. Through such regulation, an extra wind power of 475 MWh was accommodated on this day.

During the demonstration period of the first-phase project in 2017, the wind farms in the demonstration area reduced wind curtailment by 83.99 GWh, representing a decrease of 12.65% compared with that of the same period in the previous year. This is equivalent to a saving of 27,756 tons of standard coal and a reduction of roughly 73,000 tons of carbon dioxide emissions.

During the demonstration period of the second-phase project from 2020 to 2021, the daily operating cost of the heating network decreased by 4.87% compared with that of the same period in the previous year as a consequence of integrated energy management.

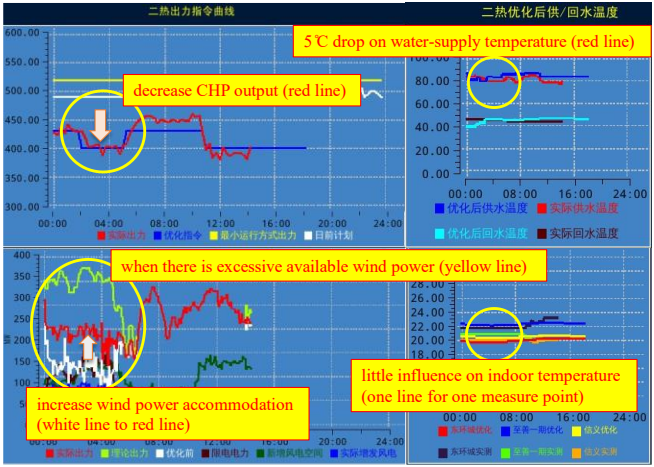

Fig. 18. A screenshot of the IEMS on March 1st, 2018.

## V. Conclusion

A novel energy circuit method for IES modeling is proposed in this article, based on which an IEMS that is able to manage large-scale IESs is further developed.

The proposed energy circuit method models natural gas networks and heating networks in the frequency domain with lump parameters, which is analogous to the electric circuit modeling of electricity networks. This method unifies the analyses of three heterogeneous energy networks in the same mathematical form. In addition, this method efficiently algebraizes the original PDE models of dynamic processes in natural gas networks and heating networks, which yields fewer variables and equations and thereby lightens the computational burden. In numerical tests, computational performance improvement across orders of magnitude is observed in both analysis and optimization. Moreover, this method also has the advantages of high parallel computing efficiency, no numerical convergence issues, and no need for initial conditions inside pipelines.

The developed IEMS is designed in a period-based manner to match the long-time-scale dynamic characteristics in IESs. This IEMS integrates various advanced applications, including dynamic state estimation, energy flow analysis, security assessment and control, and optimal energy flow, to cover the whole process of managing multiple energy flows. These advanced applications are implemented based on the ECM to enable their online uses for large-scale IESs. Since 2016, this IEMS has been deployed at 8 sites, including buildings, parks, and cities, for engineering demonstrations, and more sites with larger-scale and more comprehensive energy systems are under construction of this IEMS. The economic and environmental benefits achieved in these actual demonstration projects provide sufficient empirical evidence for the promising application prospects of IEMSs and the computing advantages of the energy circuit method.

It is worth noting that the contents of IEMSs are far more than what is presented in this article. An outlook for the future IEMS is given as follows.

1) **More energy flows**

In addition to the energy flows of electricity, natural gas, and heating introduced in this article, there are varying degrees of coupling between energy flows of steam, hydrogen, petroleum, *etc.*, and even traffic flow because of the fast development of electric vehicles. Integrating these energy flows into the managing scope of IEMSs is promising for better coordination in a wider area.

2) **More advanced applications**

The advanced applications developed in this article cover the essential but not all functions of energy flow management. More applications can be customized to satisfy diversified requirements. For instance, a virtual power plant application that aggregates distributed flexibility resources in a small-scale IES to interact with the utility grid and a nodal price calculation application that determines energy prices for different energy sectors at different positions in a large-scale IES are both useful extensions.

3) **More flexible architectures**

In this article, a centralized architecture for participants in an IES is considered, i.e., all information of different energy networks is collected together for calculation. This architecture sometimes causes privacy issues. A decentralized architecture, in which participants perform local calculations and exchange boundary information to converge to a globally consistent solution, is more flexible and practical. To realize this architecture, efficient distributed computation algorithms and robust communication networks should be further studied.

## Appendix

### A. Parameters of the Fluid Circuit and Thermal Circuit

This appendix provides energy circuit parameters of natural gas pipelines and heating pipelines, including time-domain distributed parameters, transmission parameters of two-port equivalence, and frequency-domain lump parameters.

1) **Fluid Circuit Parameters of Natural Gas Pipelines**

The TDDP fluid circuit parameters, as expressions of the original gas pipeline parameters, are as follows.

$$\begin{cases} R'_f = \lambda v_b / (AD) \\ L'_f = 1/A \\ C'_f = A/c^2 \\ k'_f = (2Dg \sin \alpha - \lambda v_b^2)/(2Dc^2) \end{cases} \quad (16)$$

Note that there is no conductance in the fluid circuit, which indicates no gas leakage under normal conditions. The controlled source parameter $k'_f$ captures the asymmetry caused by the inclination of pipelines.

The transmission parameters in the two-port equivalence of natural gas pipelines, as expressions of TDDP fluid circuit parameters, are as follows.

$$\begin{cases} \mathcal{A}_{f,\kappa} = \cosh\left(\sqrt{k_f'^2 + 4\gamma_{f,\kappa}^2}\, l/2\right)e^{-k_f' l/2} - \\ \qquad k_f' \sinh\left(\sqrt{k_f'^2 + 4\gamma_{f,\kappa}^2}\, l/2\right)e^{-k_f' l/2} / \sqrt{k_f'^2 + 4\gamma_{f,\kappa}^2} \\ \mathcal{B}_{f,\kappa} = -2\sinh\left(\sqrt{k_f'^2 + 4\gamma_{f,\kappa}^2}\, l/2\right)e^{-k_f' l/2} / \sqrt{k_f'^2 + 4/Z_{f,\kappa}^2} \\ \mathcal{C}_{f,\kappa} = -2\sinh\left(\sqrt{k_f'^2 + 4\gamma_{f,\kappa}^2}\, l/2\right)e^{-k_f' l/2} / \sqrt{k_f'^2 + 4Z_{f,\kappa}^2} \\ \mathcal{D}_{f,\kappa} = \cosh\left(\sqrt{k_f'^2 + 4\gamma_{f,\kappa}^2}\, l/2\right)e^{-k_f' l/2} + \\ \qquad k_f' \sinh\left(\sqrt{k_f'^2 + 4\gamma_{f,\kappa}^2}\, l/2\right)e^{-k_f' l/2} / \sqrt{k_f'^2 + 4\gamma_{f,\kappa}^2} \end{cases} \quad (17)$$

where $Z_{f,\kappa}$ equals $\sqrt{(R'_f + j\omega_\kappa L'_f)/j\omega_\kappa C'_f}$ and $\gamma_{f,\kappa}$ equals $\sqrt{(R'_f + j\omega_\kappa L'_f)\cdot j\omega_\kappa C'_f}$.

The FDLP fluid circuit parameters, as expressions of transmission parameters, are as follows.

$$\begin{cases} \mathcal{Z}_{f,\kappa} = -\mathcal{B}_{f,\kappa} \\ \mathcal{K}_{f,\kappa} = 1 - \mathcal{A}_{f,\kappa}\mathcal{D}_{f,\kappa} + \mathcal{B}_{f,\kappa}\mathcal{C}_{f,\kappa} \\ \mathcal{Y}_{f1,\kappa} = (\mathcal{A}_{f,\kappa}\mathcal{D}_{f,\kappa} - \mathcal{B}_{f,\kappa}\mathcal{C}_{f,\kappa} - \mathcal{A}_{f,\kappa})/\mathcal{B}_{f,\kappa} \\ \mathcal{Y}_{f2,\kappa} = (1 - \mathcal{D}_{f,\kappa})/\mathcal{B}_{f,\kappa} \end{cases} \quad (18)$$

2) **Thermal Circuit Parameters of Heating Pipelines**

The TDDP thermal circuit parameters, as expressions of the original heating pipeline parameters, are as follows.

$$\begin{cases} R'_t = \mu/(c_p^2 m^2) \\ L'_t = \rho A/(c_p m^2) \\ G'_t = \mu \\ C'_t = c_p \rho A \end{cases} \quad (19)$$

The transmission parameters in the two-port equivalence of heating pipelines, as expressions of TDDP thermal circuit parameters, are as follows.

$$\begin{cases} \mathcal{A}_{t,\kappa} = \cosh(\gamma_{t,\kappa} l) \\ \mathcal{B}_{t,\kappa} = -\sinh(\gamma_{t,\kappa} l)\cdot Z_{t,\kappa} \\ \mathcal{C}_{t,\kappa} = -\sinh(\gamma_{t,\kappa} l)/Z_{t,\kappa} \\ \mathcal{D}_{t,\kappa} = \cosh(\gamma_{t,\kappa} l) \end{cases} \quad (20)$$

where $Z_{t,\kappa}$ equals $\sqrt{(R'_t + j\omega_\kappa L'_t)/(G'_t + j\omega_\kappa C'_t)}$ and $\gamma_{t,\kappa}$ equals $\sqrt{(R'_t + j\omega_\kappa L'_t)\cdot (G'_t + j\omega_\kappa C'_t)}$.

The FDLP thermal circuit parameters, as expressions of transmission parameters, are as follows.

$$\begin{cases} \mathcal{Z}_{t,\kappa} = -\mathcal{B}_{t,\kappa} \\ \mathcal{Y}_{t,\kappa} = (1 - \mathcal{A}_{t,\kappa})/\mathcal{B}_{t,\kappa} \end{cases} \quad (21)$$

*B. Discussion about Using Historical Boundary Conditions as Surrogate Initial Conditions*

Both future boundary conditions (given values of some variables at all times) and initial conditions (given values of all variables at the initial moment) are required to determine a particular solution for the temporal-spatial PDEs of natural gas networks and heating networks. However, the ECM contains a step of discrete Fourier transform which implicitly performs a periodic extension. Consequently, initial conditions cannot be explicitly specified in this method.

To address this issue, an approach that utilizes historical boundary conditions as surrogate initial conditions is proposed, based on the intuition that the state of a physical system at an arbitrary moment does not come out of nowhere but is a result of all excitations before this moment. Considering the response decay of excitations, the historical boundary conditions with limited length are able to produce an approximate initial state close to the accurate initial state determined by the infinite-length historical boundary conditions before the initial moment, as illustrated in Fig. 19. Thus, by splicing the historical boundary conditions before the given future boundary conditions, the ECM obtains a general solution whose state at the initial moment is close to the initial conditions, i.e., the desired particular solution.

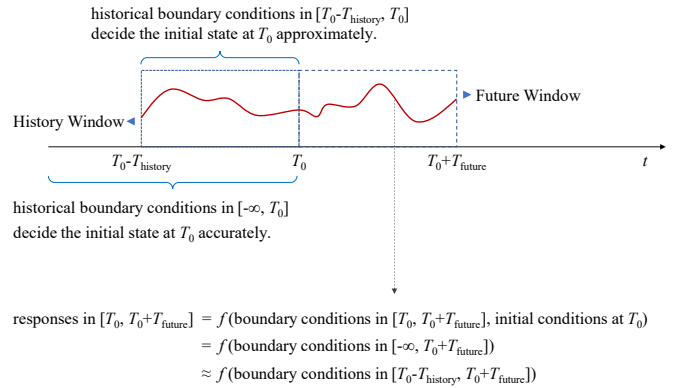

Fig. 19. Approximation of using historical boundary conditions from $T_0$-$T_{\text{history}}$ to $T_0$ as surrogate initial conditions at $T_0$.

This approximation is rather practical in engineering practice because the required historical boundary conditions are mostly the injection or potential variables of sources and loads that are adequately measured, while directly obtaining initial conditions may encounter the challenge of insufficient measurement, e.g., there are few sensors monitoring states inside natural gas pipelines and heating pipelines.

Regarding the selection of historical boundary condition length, a basic principle is that longer historical boundary conditions yield better surrogate accuracy but with a heavier computational burden at the same time. Another important lesson is that setting the total length of historical boundary conditions and future boundary conditions as an integer multiple of 24 hours can effectively decrease the error caused by the periodic extension, considering that the operation of energy networks presents day-to-day regularity. For instance, if we want to analyze the IES over the next 8 hours, then 16 hours is a good choice for the length of historical boundary conditions.

To verify the accuracy of such an approximation at different historical boundary condition lengths, energy flow analysis for the test system shown in Fig. 6 is carried out by the ECM using historical boundary conditions of 24 hours, 16 hours, 8 hours, and 0 hours as surrogate initial conditions, whose computational results of node pressure in the natural gas network and node temperature in the heating network are indicated by solid lines in Fig. 20. As a comparison, the ground truth calculated by the high-resolution FDM is indicated by the dotted lines. These numerical test results suggest that long-enough historical boundary conditions are able to accurately surrogate initial conditions. In circumstances where historical boundary conditions are inadequate or even missing, the computational results of the ECM gradually return to the ground truth over time, whose essence is that the beginning part of future boundary conditions plays the role of "historical" boundary conditions.

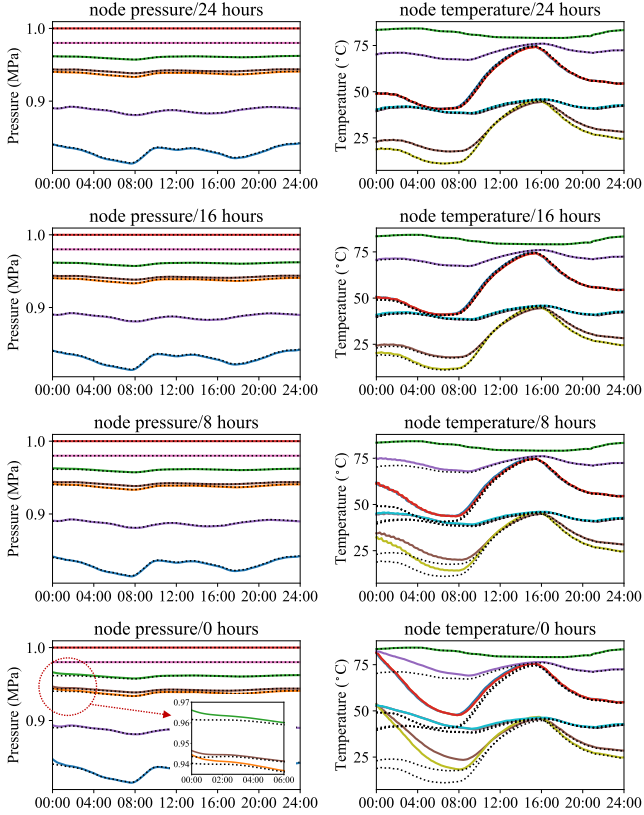

Fig. 20. Computational results of energy flow analysis: solid lines are obtained by the ECM with historical boundary conditions of different lengths, while dotted lines are obtained by the high-resolution FDM with accurate initial conditions.

*C. Mathematical Formulation of ECM-based Energy Flow Analysis*

1) **Nomenclature**

*Variables*

| | |
|---|---|
| $P_i^{(\tau)}$ | Active power injection of *i*-th bus at time $\tau$. |
| $Q_i^{(\tau)}$ | Reactive power injection of *i*-th bus at time $\tau$. |
| $V_i^{(\tau)}$ | Voltage magnitude of *i*-th bus at time $\tau$. |
| $\theta_i^{(\tau)}$ | Phase angle of *i*-th bus at time $\tau$. |
| $m_i^{(\tau)}$ | Gas flow injection of *i*-th node at time $\tau$. |
| $p_i^{(\tau)}$ | Pressure of *i*-th node at time $\tau$. |
| $h_i^{(\tau)}$ | Heat flow injection of *i*-th node at time $\tau$. |
| $T_i^{(\tau)}$ | Temperature of *i*-th node at time $\tau$. |
| $\dot{m}_i^{(\kappa)}$ | $\kappa$-th frequency component of $\{m_i^{(\tau)} \mid \tau\}$. |
| $\dot{p}_i^{(\kappa)}$ | $\kappa$-th frequency component of $\{p_i^{(\tau)} \mid \tau\}$. |
| $\dot{h}_i^{(\kappa)}$ | $\kappa$-th frequency component of $\{h_i^{(\tau)} \mid \tau\}$. |
| $\dot{T}_i^{(\kappa)}$ | $\kappa$-th frequency component of $\{T_i^{(\tau)} \mid \tau\}$. |

*Parameters*

| | |
|---|---|
| $G_{ij} + jB_{ij}$ | Entry at *i*-th row and *j*-th column in the node admittance matrix of electricity networks. |
| $Y_{f,ij}^{(\kappa)}$ | Entry at *i*-th row and *j*-th column in the $\kappa$-th node admittance matrix of natural gas networks. |
| $Y_{t,ij}^{(\kappa)}$ | Entry at *i*-th row and *j*-th column in the $\kappa$-th node admittance matrix of heating networks. |
| $c_{e,ij}$ | Coupling coefficients of *i*-th energy conversion device to active power injection of *j*-th bus. |
| $c_{f,ij}$ | Coupling coefficients of *i*-th energy conversion device to gas flow injection of *j*-th node. |
| $c_{t,ij}$ | Coupling coefficients of *i*-th energy conversion device to heat flow injection of *j*-th node. |
| $N$ | Number of time-domain calculation points. |

2) **Formulation**

The ECM-based energy flow analysis model includes the following algebraic equations.

(i) AC power flow equations of electricity networks

$$\begin{cases} P_i^{(\tau)} = V_i^{(\tau)} \sum_j V_j^{(\tau)} [G_{ij} \cos(\theta_i^{(\tau)} - \theta_j^{(\tau)}) + B_{ij} \sin(\theta_i^{(\tau)} - \theta_j^{(\tau)})] \\ Q_i^{(\tau)} = V_i^{(\tau)} \sum_j V_j^{(\tau)} [G_{ij} \sin(\theta_i^{(\tau)} - \theta_j^{(\tau)}) - B_{ij} \cos(\theta_i^{(\tau)} - \theta_j^{(\tau)})] \end{cases} \quad (22)$$

(ii) fluid circuit equations of natural gas networks

$$\dot{m}_i^{(\kappa)} = \sum_j Y_{f,ij}^{(\kappa)} \dot{p}_j^{(\kappa)} \quad (23)$$

(iii) thermal circuit equations of heating networks

$$\dot{h}_i^{(\kappa)} = \sum_j Y_{t,ij}^{(\kappa)} \dot{T}_j^{(\kappa)} \quad (24)$$

(iv) energy conversion equations

Due to the dynamics of coupling devices are much faster than that of natural gas networks and heating networks, their transition processes between two adjacent states are ignorable so that a steady-state coupling model is adopted as follows.

$$\sum_j c_{e,ij} P_j^{(\tau)} + \sum_j c_{f,ij} m_j^{(\tau)} + \sum_j c_{t,ij} h_j^{(\tau)} = 0 \quad (25)$$

(v) time-frequency transformation equations

$$\begin{cases} \sum_\kappa \mathrm{Re}(\dot{m}_i^{(\kappa)} e^{j2\pi\kappa\tau/N}) = m_i^{(\tau)} \\ \sum_\kappa \mathrm{Re}(\dot{p}_i^{(\kappa)} e^{j2\pi\kappa\tau/N}) = p_i^{(\tau)} \\ \sum_\kappa \mathrm{Re}(\dot{h}_i^{(\kappa)} e^{j2\pi\kappa\tau/N}) = h_i^{(\tau)} \\ \sum_\kappa \mathrm{Re}(\dot{T}_i^{(\kappa)} e^{j2\pi\kappa\tau/N}) = T_i^{(\tau)} \end{cases} \quad (26)$$

which is equivalent to

$$\begin{cases} \sum_\kappa \mathrm{Re}(\dot{m}_i^{(\kappa)})\cos(\frac{2\pi\kappa\tau}{N}) - \mathrm{Im}(\dot{m}_i^{(\kappa)})\sin(\frac{2\pi\kappa\tau}{N}) = m_i^{(\tau)} \\ \sum_\kappa \mathrm{Re}(\dot{p}_i^{(\kappa)})\cos(\frac{2\pi\kappa\tau}{N}) - \mathrm{Im}(\dot{p}_i^{(\kappa)})\sin(\frac{2\pi\kappa\tau}{N}) = p_i^{(\tau)} \\ \sum_\kappa \mathrm{Re}(\dot{h}_i^{(\kappa)})\cos(\frac{2\pi\kappa\tau}{N}) - \mathrm{Im}(\dot{h}_i^{(\kappa)})\sin(\frac{2\pi\kappa\tau}{N}) = h_i^{(\tau)} \\ \sum_\kappa \mathrm{Re}(\dot{T}_i^{(\kappa)})\cos(\frac{2\pi\kappa\tau}{N}) - \mathrm{Im}(\dot{T}_i^{(\kappa)})\sin(\frac{2\pi\kappa\tau}{N}) = T_i^{(\tau)} \end{cases} \quad (27)$$

*D. Mathematical Formulation of ECM-based Optimal Energy Flow*

1) **Nomenclature**

*Variables*

(1) time-domain controlled variables

$P_{s,i}^{(\tau)}$ — Controllable power production of $i$-th bus at time $\tau$.

$m_{s,i}^{(\tau)}$ — Controllable gas production of $i$-th node at time $\tau$.

$h_{s,i}^{(\tau)}$ — Controllable heat production of $i$-th node at time $\tau$.

$P_{c,i}^{(\tau)}$ — Controllable power load of $i$-th bus at time $\tau$.

$m_{c,i}^{(\tau)}$ — Controllable gas load of $i$-th node at time $\tau$.

$h_{c,i}^{(\tau)}$ — Controllable heat load of $i$-th node at time $\tau$.

$P_{cg,i}^{(\tau)}$ — Charging power of the storage device of $i$-th node at time $\tau$.

$m_{cg,i}^{(\tau)}$ — Storing natural gas of the storage device of $i$-th node at time $\tau$.

$h_{cg,i}^{(\tau)}$ — Storing thermal power of the storage device of $i$-th node at time $\tau$.

$P_{dc,i}^{(\tau)}$ — Discharging power of the storage device of $i$-th node at time $\tau$.

$m_{dc,i}^{(\tau)}$ — Releasing natural gas of the storage device of $i$-th node at time $\tau$.

$h_{dc,i}^{(\tau)}$ — Releasing thermal power of the storage device of $i$-th node at time $\tau$.

$E_{e,i}^{(\tau)}$ — Stored electrical energy in the storage device of $i$-th node at time $\tau$.

$E_{f,i}^{(\tau)}$ — Stored natural gas in the storage device of $i$-th node at time $\tau$.

$E_{t,i}^{(\tau)}$ — Stored thermal energy in the storage device of $i$-th node at time $\tau$.

(2) time-domain monitored variables

$P_{L,i}^{(\tau)}$ — Power of $i$-th transmission line at time $\tau$.

$p_i^{(\tau)}$ — Pressure of $i$-th node at time $\tau$.

$T_i^{(\tau)}$ — Temperature of $i$-th node at time $\tau$.

(3) frequency-domain controlled variables

$\dot{m}_{s,i}^{(\kappa)}$ — $\kappa$-th frequency component of $\{m_{s,i}^{(\tau)} \mid \tau\}$.

$\dot{h}_{s,i}^{(\kappa)}$ — $\kappa$-th frequency component of $\{h_{s,i}^{(\tau)} \mid \tau\}$.

$\dot{m}_{c,i}^{(\kappa)}$ — $\kappa$-th frequency component of $\{m_{c,i}^{(\tau)} \mid \tau\}$.

$\dot{h}_{c,i}^{(\kappa)}$ — $\kappa$-th frequency component of $\{h_{c,i}^{(\tau)} \mid \tau\}$.

$\dot{m}_{cg,i}^{(\kappa)}$ — $\kappa$-th frequency component of $\{m_{cg,i}^{(\tau)} \mid \tau\}$.

$\dot{h}_{cg,i}^{(\kappa)}$ — $\kappa$-th frequency component of $\{h_{cg,i}^{(\tau)} \mid \tau\}$.

$\dot{m}_{dc,i}^{(\kappa)}$ — $\kappa$-th frequency component of $\{m_{dc,i}^{(\tau)} \mid \tau\}$.

$\dot{h}_{dc,i}^{(\kappa)}$ — $\kappa$-th frequency component of $\{h_{dc,i}^{(\tau)} \mid \tau\}$.

(4) frequency-domain monitored variables

$\dot{p}_i^{(\kappa)}$ — $\kappa$-th frequency component of $\{p_i^{(\tau)} \mid \tau\}$.

$\dot{T}_i^{(\kappa)}$ — $\kappa$-th frequency component of $\{T_i^{(\tau)} \mid \tau\}$.

*Parameters*

$\underline{P}_{s,i}, \overline{P}_{s,i}$ — Lower and upper bounds of $P_{s,i}^{(\tau)}$.

$\underline{m}_{s,i}, \overline{m}_{s,i}$ — Lower and upper bounds of $m_{s,i}^{(\tau)}$.

$\underline{h}_{s,i}, \overline{h}_{s,i}$ — Lower and upper bounds of $h_{s,i}^{(\tau)}$.

$\underline{P}_{c,i}, \overline{P}_{c,i}$ — Lower and upper bounds of $P_{c,i}^{(\tau)}$.

$\underline{m}_{c,i}, \overline{m}_{c,i}$ — Lower and upper bounds of $m_{c,i}^{(\tau)}$.

$\underline{h}_{c,i}, \overline{h}_{c,i}$ — Lower and upper bounds of $h_{c,i}^{(\tau)}$.

$\underline{P}_{L,i}, \overline{P}_{L,i}$ — Lower and upper bounds of $P_{L,i}^{(\tau)}$.

$\underline{p}_i, \overline{p}_i$ — Lower and upper bounds of $p_i^{(\tau)}$.

$\underline{T}_i, \overline{T}_i$ — Lower and upper bounds of $T_i^{(\tau)}$.

$\underline{P}_{cg,i}, \overline{P}_{cg,i}$ — Lower and upper bounds of $P_{cg,i}^{(\tau)}$.

$\underline{m}_{cg,i}, \overline{m}_{cg,i}$ — Lower and upper bounds of $m_{cg,i}^{(\tau)}$.

$\underline{h}_{cg,i}, \overline{h}_{cg,i}$ — Lower and upper bounds of $h_{cg,i}^{(\tau)}$.

$\underline{P}_{dc,i}, \overline{P}_{dc,i}$ — Lower and upper bounds of $P_{dc,i}^{(\tau)}$.

$\underline{m}_{dc,i}, \overline{m}_{dc,i}$ — Lower and upper bounds of $m_{dc,i}^{(\tau)}$.

$\underline{h}_{dc,i}, \overline{h}_{dc,i}$ — Lower and upper bounds of $h_{dc,i}^{(\tau)}$.

$\underline{E}_{e,i}, \overline{E}_{e,i}$ — Lower and upper bounds of $E_{e,i}^{(\tau)}$.

$\underline{E}_{f,i}, \overline{E}_{f,i}$ — Lower and upper bounds of $E_{f,i}^{(\tau)}$.

$\underline{E}_{t,i}, \overline{E}_{t,i}$ — Lower and upper bounds of $E_{t,i}^{(\tau)}$.

$\underline{P}_{\Delta s,i}, \overline{P}_{\Delta s,i}$ — Down and up ramping limits of $P_{s,i}^{(\tau)}$.

$\underline{m}_{\Delta s,i}, \overline{m}_{\Delta s,i}$ — Down and up ramping limits of $m_{s,i}^{(\tau)}$.

| | |
|---|---|
| $\underline{h}_{\Delta s,i}$, $\overline{h}_{\Delta s,i}$ | Down and up ramping limits of $h_{s,i}^{(\tau)}$. |
| $P_{d,i}^{(\tau)}$ | Fixed power load of $i$-th bus at time $\tau$. |
| $m_{d,i}^{(\tau)}$ | Fixed gas load of $i$-th node at time $\tau$. |
| $h_{d,i}^{(\tau)}$ | Fixed heat load of $i$-th node at time $\tau$. |
| $\dot{m}_{d,i}^{(\kappa)}$ | $\kappa$-th frequency component of $\{\dot{m}_{d,i}^{(\kappa)} \mid \tau\}$. |
| $\dot{h}_{d,i}^{(\kappa)}$ | $\kappa$-th frequency component of $\{\dot{h}_{d,i}^{(\kappa)} \mid \tau\}$. |
| $S_{ij}$ | Entry at $i$-th row and $j$-th column in the power transformation distribution factor matrix of electricity networks. |
| $Y_{f,ij}^{(\kappa)}$ | Entry at $i$-th row and $j$-th column in the $\kappa$-th node admittance matrix of gas networks. |
| $Y_{t,ij}^{(\kappa)}$ | Entry at $i$-th row and $j$-th column in the $\kappa$-th node admittance matrix of heating networks. |
| $c_{1,ij} \sim c_{6,ij}$ | Coupling coefficients of $i$-th energy conversion device to injection of $j$-th bus/node. |
| $u_{0,e,i} \sim u_{2,e,i}$ | Cost coefficients of $P_{s,i}^{(\tau)}$. |
| $u_{0,f,i} \sim u_{2,f,i}$ | Cost coefficients of $m_{s,i}^{(\tau)}$. |
| $u_{0,t,i} \sim u_{2,t,i}$ | Cost coefficients of $h_{s,i}^{(\tau)}$. |
| $\eta_{e,cg,i}$, $\eta_{e,dc,i}$ | Charging and discharging efficiency of the electrical energy storage device at $i$-th node. |
| $\eta_{f,cg,i}$, $\eta_{f,dc,i}$ | Storing and releasing efficiency of the natural gas storage device at $i$-th node. |
| $\eta_{t,cg,i}$, $\eta_{t,dc,i}$ | Storing and releasing efficiency of the thermal energy storage device at $i$-th node. |
| $\Delta T$ | Length of each time step. |
| $N$ | Number of time-domain calculation points. |

2) **Formulation**

The objective function of optimal energy flow is usually minimizing the total operating cost that is quadratic or linear to the production of controllable devices, as formulated in (28).

$$\min \sum_{\tau} \left( \begin{array}{l} \sum_{i} u_{2,e,i}(P_{s,i}^{(\tau)})^2 + u_{1,e,i}P_{s,i}^{(\tau)} + u_{0,e,i} + \\ \sum_{i} u_{2,f,i}(m_{s,i}^{(\tau)})^2 + u_{1,f,i}m_{s,i}^{(\tau)} + u_{0,f,i} + \\ \sum_{i} u_{2,t,i}(h_{s,i}^{(\tau)})^2 + u_{1,t,i}h_{s,i}^{(\tau)} + u_{0,t,i} \end{array} \right) \quad (28)$$

The constraints of ECM-based optimal energy flow are as follows, in which (i)-(v) are operation constraints for time-domain controlled variables, (vi) is security constraints for time-domain monitored variables, (vii) is network constraints that couple controlled variables and monitored variables, and (viii) is time-frequency transformation constraints that couple time-domain variables and frequency-domain variables.

(i) output ranges of devices

$$\begin{cases} \underline{P}_{s,i} \leq P_{s,i}^{(\tau)} \leq \overline{P}_{s,i} & \underline{P}_{c,i} \leq P_{c,i}^{(\tau)} \leq \overline{P}_{c,i} \\ \underline{P}_{cg,i} \leq P_{cg,i}^{(\tau)} \leq \overline{P}_{cg,i} & \underline{P}_{dc,i} \leq P_{dc,i}^{(\tau)} \leq \overline{P}_{dc,i} \\ \underline{m}_{s,i} \leq m_{s,i}^{(\tau)} \leq \overline{m}_{s,i} & \underline{m}_{c,i} \leq m_{c,i}^{(\tau)} \leq \overline{m}_{c,i} \\ \underline{m}_{cg,i} \leq m_{cg,i}^{(\tau)} \leq \overline{m}_{cg,i} & \underline{m}_{dc,i} \leq m_{dc,i}^{(\tau)} \leq \overline{m}_{dc,i} \\ \underline{h}_{s,i} \leq h_{s,i}^{(\tau)} \leq \overline{h}_{s,i} & \underline{h}_{c,i} \leq h_{c,i}^{(\tau)} \leq \overline{h}_{c,i} \\ \underline{h}_{cg,i} \leq h_{cg,i}^{(\tau)} \leq \overline{h}_{cg,i} & \underline{h}_{dc,i} \leq h_{dc,i}^{(\tau)} \leq \overline{h}_{dc,i} \end{cases} \quad (29)$$

(ii) ramping limits of devices

$$\begin{cases} -\underline{P}_{\Delta s,i} \leq P_{s,i}^{(\tau)} - P_{s,i}^{(\tau-1)} \leq \overline{P}_{\Delta s,i} \\ -\underline{m}_{\Delta s,i} \leq m_{s,i}^{(\tau)} - m_{s,i}^{(\tau-1)} \leq \overline{m}_{\Delta s,i} \\ -\underline{h}_{\Delta s,i} \leq h_{s,i}^{(\tau)} - h_{s,i}^{(\tau-1)} \leq \overline{h}_{\Delta s,i} \end{cases} \quad (30)$$

(iii) energy conversion relationships of devices

An energy-hub model is adopted to describe the energy conversion relationships as follows.

$$\begin{array}{l} \sum_{j} c_{1,ij} P_{s,j}^{(\tau)} + \sum_{j} c_{2,ij} m_{s,j}^{(\tau)} + \sum_{j} c_{3,ij} h_{s,j}^{(\tau)} + \\ \sum_{j} c_{4,ij} P_{c,j}^{(\tau)} + \sum_{j} c_{5,ij} m_{c,j}^{(\tau)} + \sum_{j} c_{6,ij} h_{c,j}^{(\tau)} \end{array} = 0 \quad (31)$$

(iv) real-time power balance of electricity networks

$$\sum_{i} (P_{s,i}^{(\tau)} + P_{dc,i}^{(\tau)} - P_{c,i}^{(\tau)} - P_{d,i}^{(\tau)} - P_{cg,i}^{(\tau)}) = 0 \quad (32)$$

(v) energy storage constraints [114]-[116]

$$\begin{cases} E_{e,i}^{(\tau)} = E_{e,i}^{(\tau-1)} + (\eta_{e,cg,i} P_{cg,i}^{(\tau)} - P_{dc,i}^{(\tau)} / \eta_{e,dc,i}) \Delta T \\ E_{f,i}^{(\tau)} = E_{f,i}^{(\tau-1)} + (\eta_{f,cg,i} m_{cg,i}^{(\tau)} - m_{dc,i}^{(\tau)} / \eta_{f,dc,i}) \Delta T \\ E_{t,i}^{(\tau)} = E_{t,i}^{(\tau-1)} + (\eta_{t,cg,i} h_{cg,i}^{(\tau)} - h_{dc,i}^{(\tau)} / \eta_{t,dc,i}) \Delta T \\ \underline{E}_{e,i} \leq E_{e,i}^{(\tau)} \leq \overline{E}_{e,i} \\ \underline{E}_{f,i} \leq E_{f,i}^{(\tau)} \leq \overline{E}_{f,i} \\ \underline{E}_{t,i} \leq E_{t,i}^{(\tau)} \leq \overline{E}_{t,i} \end{cases} \quad (33)$$

(vi) security bounds of energy networks

$$\begin{cases} \underline{P}_{L,i} \leq P_{L,i}^{(\tau)} \leq \overline{P}_{L,i} \\ \underline{p}_i \leq p_i^{(\tau)} \leq \overline{p}_i \\ \underline{T}_i \leq T_i^{(\tau)} \leq \overline{T}_i \end{cases} \quad (34)$$

(vii) network constraints

$$P_{L,i}^{(\tau)} = \sum_{j} S_{ij} (P_{s,j}^{(\tau)} + P_{dc,j}^{(\tau)} - P_{c,j}^{(\tau)} - P_{d,j}^{(\tau)} - P_{cg,j}^{(\tau)}) \quad (35)$$

$$\dot{m}_{s,i}^{(\kappa)} + \dot{m}_{dc,i}^{(\kappa)} - \dot{m}_{c,i}^{(\kappa)} - \dot{m}_{d,i}^{(\kappa)} - \dot{m}_{cg,i}^{(\kappa)} = \sum_{j} Y_{f,ij}^{(\kappa)} \dot{p}_j^{(\kappa)} \quad (36)$$

$$\dot{h}_{s,i}^{(\kappa)} + \dot{h}_{dc,i}^{(\kappa)} - \dot{h}_{c,i}^{(\kappa)} - \dot{h}_{d,i}^{(\kappa)} - \dot{h}_{cg,i}^{(\kappa)} = \sum_{j} Y_{t,ij}^{(\kappa)} \dot{T}_j^{(\kappa)} \quad (37)$$

where, (35) is the DC power flow model of electricity networks, (36) is the fluid circuit model of natural gas networks, and (37) is the thermal circuit model of heating networks.

(viii) time-frequency transformation constraints

$$\begin{cases} \sum_\kappa \text{Re}(\dot{m}_{s,i}^{(\kappa)})\cos(\frac{2\pi\kappa\tau}{N}) - \text{Im}(\dot{m}_{s,i}^{(\kappa)})\sin(\frac{2\pi\kappa\tau}{N}) = m_{s,i}^{(\tau)} \\ \sum_\kappa \text{Re}(\dot{m}_{c,i}^{(\kappa)})\cos(\frac{2\pi\kappa\tau}{N}) - \text{Im}(\dot{m}_{c,i}^{(\kappa)})\sin(\frac{2\pi\kappa\tau}{N}) = m_{c,i}^{(\tau)} \\ \sum_\kappa \text{Re}(\dot{m}_{cg,i}^{(\kappa)})\cos(\frac{2\pi\kappa\tau}{N}) - \text{Im}(\dot{m}_{cg,i}^{(\kappa)})\sin(\frac{2\pi\kappa\tau}{N}) = m_{cg,i}^{(\tau)} \\ \sum_\kappa \text{Re}(\dot{m}_{dc,i}^{(\kappa)})\cos(\frac{2\pi\kappa\tau}{N}) - \text{Im}(\dot{m}_{dc,i}^{(\kappa)})\sin(\frac{2\pi\kappa\tau}{N}) = m_{dc,i}^{(\tau)} \\ \sum_\kappa \text{Re}(\dot{h}_{s,i}^{(\kappa)})\cos(\frac{2\pi\kappa\tau}{N}) - \text{Im}(\dot{h}_{s,i}^{(\kappa)})\sin(\frac{2\pi\kappa\tau}{N}) = h_{s,i}^{(\tau)} \\ \sum_\kappa \text{Re}(\dot{h}_{c,i}^{(\kappa)})\cos(\frac{2\pi\kappa\tau}{N}) - \text{Im}(\dot{h}_{c,i}^{(\kappa)})\sin(\frac{2\pi\kappa\tau}{N}) = h_{c,i}^{(\tau)} \\ \sum_\kappa \text{Re}(\dot{h}_{cg,i}^{(\kappa)})\cos(\frac{2\pi\kappa\tau}{N}) - \text{Im}(\dot{h}_{cg,i}^{(\kappa)})\sin(\frac{2\pi\kappa\tau}{N}) = h_{cg,i}^{(\tau)} \\ \sum_\kappa \text{Re}(\dot{h}_{dc,i}^{(\kappa)})\cos(\frac{2\pi\kappa\tau}{N}) - \text{Im}(\dot{h}_{dc,i}^{(\kappa)})\sin(\frac{2\pi\kappa\tau}{N}) = h_{dc,i}^{(\tau)} \\ \sum_\kappa \text{Re}(\dot{p}_i^{(\kappa)})\cos(\frac{2\pi\kappa\tau}{N}) - \text{Im}(\dot{p}_i^{(\kappa)})\sin(\frac{2\pi\kappa\tau}{N}) = p_i^{(\tau)} \\ \sum_\kappa \text{Re}(\dot{T}_i^{(\kappa)})\cos(\frac{2\pi\kappa\tau}{N}) - \text{Im}(\dot{T}_i^{(\kappa)})\sin(\frac{2\pi\kappa\tau}{N}) = T_i^{(\tau)} \end{cases} \quad (38)$$

Note that the above constraints are linear (convex) because the trigonometric terms are all constant coefficients.

*E. Mathematical Formulation of ECM-based Dynamic State Estimation*

1) **Nomenclature**

*Variables*

| | |
|---|---|
| $P_i^{(\tau)}$ | Estimated active power injection of *i*-th bus at time $\tau$. |
| $Q_i^{(\tau)}$ | Estimated reactive power injection of *i*-th bus at time $\tau$. |
| $V_i^{(\tau)}$ | Estimated voltage magnitude of *i*-th bus at time $\tau$. |
| $\theta_i^{(\tau)}$ | Estimated phase angle of *i*-th bus at time $\tau$. |
| $m_i^{(\tau)}$ | Estimated gas flow injection of *i*-th node at time $\tau$. |
| $p_i^{(\tau)}$ | Estimated pressure of *i*-th node at time $\tau$. |
| $h_i^{(\tau)}$ | Estimated heat flow injection of *i*-th node at time $\tau$. |
| $T_i^{(\tau)}$ | Estimated temperature of *i*-th node at time $\tau$. |
| $\dot{m}_i^{(\kappa)}$ | $\kappa$-th frequency component of $\{m_i^{(\tau)} \mid \tau\}$. |
| $\dot{p}_i^{(\kappa)}$ | $\kappa$-th frequency component of $\{p_i^{(\tau)} \mid \tau\}$. |
| $\dot{h}_i^{(\kappa)}$ | $\kappa$-th frequency component of $\{h_i^{(\tau)} \mid \tau\}$. |
| $\dot{T}_i^{(\kappa)}$ | $\kappa$-th frequency component of $\{T_i^{(\tau)} \mid \tau\}$. |

*Parameters*

| | |
|---|---|
| $\hat{P}_i^{(\tau)}$ | Measured active power injection of *i*-th bus at time $\tau$. |
| $\hat{Q}_i^{(\tau)}$ | Measured reactive power injection of *i*-th bus at time $\tau$. |
| $\hat{V}_i^{(\tau)}$ | Measured voltage magnitude of *i*-th bus at time $\tau$. |
| $\hat{m}_i^{(\tau)}$ | Measured gas flow injection of *i*-th node at time $\tau$. |
| $\hat{p}_i^{(\tau)}$ | Measured pressure of *i*-th node at time $\tau$. |
| $\hat{h}_i^{(\tau)}$ | Measured heat flow injection of *i*-th node at time $\tau$. |
| $\hat{T}_i^{(\tau)}$ | Measured temperature of *i*-th node at time $\tau$. |
| $G_{ij} + jB_{ij}$ | Entry at *i*-th row and *j*-th column in the node admittance matrix of electricity networks. |
| $Y_{f,ij}^{(\kappa)}$ | Entry at *i*-th row and *j*-th column in the $\kappa$-th node admittance matrix of natural gas networks. |
| $Y_{t,ij}^{(\kappa)}$ | Entry at *i*-th row and *j*-th column in the $\kappa$-th node admittance matrix of heating networks. |
| $c_{e,ij}$ | Coupling coefficients of *i*-th energy conversion device to active power injection of *j*-th bus. |
| $c_{f,ij}$ | Coupling coefficients of *i*-th energy conversion device to gas flow injection of *j*-th node. |
| $c_{t,ij}$ | Coupling coefficients of *i*-th energy conversion device to heat flow injection of *j*-th node. |
| $\delta^{(\tau)}$ | Weight of measurement residuals at time $\tau$. |
| $N$ | Number of time-domain calculation points. |

2) **Formulation**

The objective function of dynamic state estimation is usually minimizing the weighted sum of measurement residuals, as formulated in (39).

$$\min \sum_\tau \delta^{(\tau)} \begin{pmatrix} \sum_{i\in\Omega_P}(P_i^{(\tau)}-\hat{P}_i^{(\tau)})^2 + \sum_{i\in\Omega_Q}(Q_i^{(\tau)}-\hat{Q}_i^{(\tau)})^2 + \\ \sum_{i\in\Omega_V}(V_i^{(\tau)}-\hat{V}_i^{(\tau)})^2 + \sum_{i\in\Omega_m}(m_i^{(\tau)}-\hat{m}_i^{(\tau)})^2 + \\ \sum_{i\in\Omega_p}(p_i^{(\tau)}-\hat{p}_i^{(\tau)})^2 + \sum_{i\in\Omega_h}(h_i^{(\tau)}-\hat{h}_i^{(\tau)})^2 + \\ \sum_{i\in\Omega_T}(T_i^{(\tau)}-\hat{T}_i^{(\tau)})^2 \end{pmatrix} \quad (39)$$

where, $\Omega_P$, $\Omega_Q$, $\Omega_V$, $\Omega_m$, $\Omega_p$, $\Omega_h$, and $\Omega_T$ distinguish the buses/nodes that are with measurement of active power, reactive power, voltage magnitude, gas flow, pressure, heat flow, and temperature.

The constraints of dynamic state estimation are as follows:

(i) AC power flow model of electricity networks

$$\begin{cases} P_i^{(\tau)} = V_i^{(\tau)}\sum_j V_j^{(\tau)}[G_{ij}\cos(\theta_i^{(\tau)}-\theta_j^{(\tau)}) + B_{ij}\sin(\theta_i^{(\tau)}-\theta_j^{(\tau)})] \\ Q_i^{(\tau)} = V_i^{(\tau)}\sum_j V_j^{(\tau)}[G_{ij}\sin(\theta_i^{(\tau)}-\theta_j^{(\tau)}) - B_{ij}\cos(\theta_i^{(\tau)}-\theta_j^{(\tau)})] \end{cases} \quad (40)$$

(ii) fluid circuit model of natural gas networks

$$\dot{m}_i^{(\kappa)} = \sum_j Y_{f,ij}^{(\kappa)} \dot{p}_j^{(\kappa)} \quad (41)$$

(iii) thermal circuit model of heating networks

$$\dot{h}_i^{(\kappa)} = \sum_j Y_{t,ij}^{(\kappa)} \dot{T}_j^{(\kappa)} \quad (42)$$

(iv) time-frequency transformation constraints

$$\begin{cases} \sum_\kappa \text{Re}(\dot{m}_i^{(\kappa)})\cos(\frac{2\pi\kappa\tau}{N}) - \text{Im}(\dot{m}_i^{(\kappa)})\sin(\frac{2\pi\kappa\tau}{N}) = m_i^{(\tau)} \\ \sum_\kappa \text{Re}(\dot{p}_i^{(\kappa)})\cos(\frac{2\pi\kappa\tau}{N}) - \text{Im}(\dot{p}_i^{(\kappa)})\sin(\frac{2\pi\kappa\tau}{N}) = p_i^{(\tau)} \\ \sum_\kappa \text{Re}(\dot{h}_i^{(\kappa)})\cos(\frac{2\pi\kappa\tau}{N}) - \text{Im}(\dot{h}_i^{(\kappa)})\sin(\frac{2\pi\kappa\tau}{N}) = h_i^{(\tau)} \\ \sum_\kappa \text{Re}(\dot{T}_i^{(\kappa)})\cos(\frac{2\pi\kappa\tau}{N}) - \text{Im}(\dot{T}_i^{(\kappa)})\sin(\frac{2\pi\kappa\tau}{N}) = T_i^{(\tau)} \end{cases} \quad (43)$$

(v) energy coupling constraints

$$\sum_j c_{e,ij} P_j^{(\tau)} + \sum_j c_{f,ij} m_j^{(\tau)} + \sum_j c_{t,ij} h_j^{(\tau)} = 0 \quad (44)$$

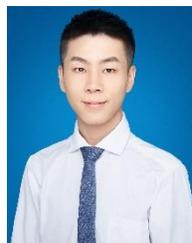

**Binbin Chen** (Student Member, IEEE) received the B.S. degree from the Department of Electrical Engineering, Zhejiang University, Zhejiang, China, in 2018. He is currently working toward the Ph.D. degree in the Department of Electrical Engineering, Tsinghua University. His research interest focuses on the energy management of integrated energy systems.

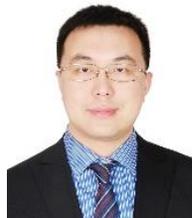

**Qinglai Guo** (Senior Member, IEEE) was born in Jilin City, Jilin Province in China on March 6, 1979. He received the B.S. degree from the Department of Electrical Engineering, Tsinghua University, Beijing, China, in 2000, and the Ph.D. degree from Tsinghua University in 2005. He is currently a Professor with Tsinghua University. His research interests include energy management systems, voltage stability and control, and cyber-physical systems. He is currently an IET Fellow and a CIGRE Member, and is involved in five workgroups of these organizations. He is the TCPC of the Energy Internet Coordinating Committee of IEEE PES, the Co-Chair of IEEE PES Work Group on Energy Internet, IEEE PES Task Force on Cyber-Physical Interdependence for Power System Operation and Control, and IEEE PES Task Force on Voltage Control for Smart Grid. He is the Editorial Member of the IEEE TRANSACTIONS ON POWER SYSTEMS, RENEWABLE AND SUSTAINABLE ENERGY REVIEWS, and IEEE TRANSACTIONS ON SMART GRID.

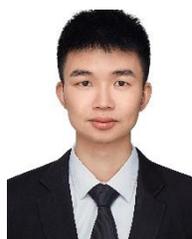

**Guanxiong Yin** (Student Member, IEEE) received the B.S. degree from the Department of Electrical Engineering, Tsinghua University, Beijing, China, in 2018. He is currently working toward the Ph.D. degree with the Department of Electrical Engineering, Tsinghua University. His research interests include the coordinated operation, state estimation, and flexibility utilization in integrated energy systems.

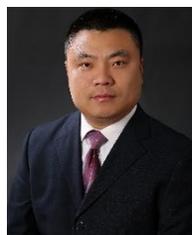

**Bin Wang** received the B.S. and Ph.D. degrees from the Department of Electrical Engineering, Tsinghua University, Beijing, China, in 2005 and 2011, respectively. He is currently a Research Scientist with the Department of Electrical Engineering, Tsinghua University. His research interests include renewable energy optimal dispatch and control, and automatic voltage control.

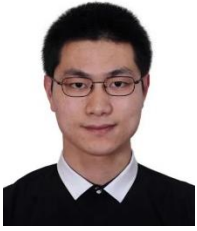

**Zhaoguang Pan** (Member IEEE) received the B.S. and Ph.D. degrees from the Department of Electrical Engineering, Tsinghua University, Beijing, China, in 2013 and 2018, respectively. His research interests include integrated energy systems, energy management, and energy Internet.

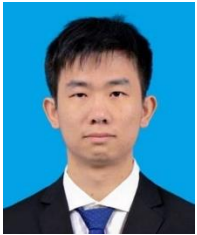

**Yuwei Chen** received the B.S. and Ph. D. degrees from the Department of Electrical Engineering, Tsinghua University, Beijing, China, in 2016 and 2021, respectively. His research interests include energy management and microgrid analysis and operation.

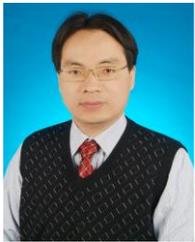

**Wenchuan Wu** (Fellow, IEEE) received the B.S., M.S., and Ph.D. degrees from the Department of Electrical Engineering, Tsinghua University, Beijing, China. He is currently a Professor with the Department of Electrical Engineering, Tsinghua University. He is currently a Fellow of IEEE and IET, and an Associate Editor of *IET Generation, Transmission and Distribution* and *IET Energy Systems Integration*. His research interests include energy management system, active distribution system operation and control, and machine learning and its application in energy system.

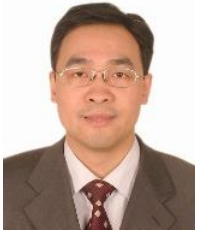

**Hongbin Sun** (Fellow, IEEE) received his double B.S. degrees from Tsinghua University in 1992, the Ph.D. from Department of Electrical Engineering, Tsinghua University in 1997. He is now Changjiang Scholar Chair Professor of Education Ministry of China, tenured full professor of electrical engineering and the director of Energy Management and Control Research Center in Tsinghua University. From 2007.9 to 2008.9, he was a visiting professor with School of EECS at the Washington State University. He is a Fellow of IEEE and IET. He is serving as a chair of IEEE smart grid voltage control task force and IEEE Energy Internet working group. He served as the founding Chair of IEEE Conference on Energy Internet and Energy System Integration in Nov 2017. In recent 20 years, he led a research group in Tsinghua University to develop a commercial system-wide automatic voltage control system, which has been applied to more than 100 electrical power control centers in China as well as the control center of PJM interconnection, the largest regional power grid in USA. He published more than 300 peer-reviewed papers. He won the China National Technology Innovation Award in 2008, the National Distinguished Teacher Award in China in 2009, and the National Science Fund for Distinguished Young Scholars of China in 2010. His research interests include energy management system, automatic voltage control, Energy Internet and energy system integration.